\documentclass[sn-mathphys,Numbered]{sn-jnl}

\usepackage{graphicx}%
\usepackage{multirow}%
\usepackage{amsmath,amssymb,amsfonts}%
\usepackage{amsthm}%
\usepackage{mathrsfs}%
\usepackage[title]{appendix}%
\usepackage{xcolor}%
\usepackage{textcomp}%
\usepackage{manyfoot}%
\usepackage{booktabs}%
\usepackage{algorithm}%
\usepackage{algorithmicx}%
\usepackage{algpseudocode}%
\usepackage{listings}%
\usepackage{adjustbox}

\usepackage{ulem}

\newcommand{\eg}{\textit{e.g.}}
\newcommand{\ea}{\textit{et al.}}

\newcommand{\new}[1]{\textcolor{black}{#1}}
\usepackage{ulem}

\begin{document}

\title[The dynamics of the Reddit collective action leading to the GameStop short squeeze]{The dynamics of the Reddit collective action leading to the GameStop short squeeze}

\author*[a,b,c]{\fnm{Antonio} \sur{Desiderio}}
\author[d,e]{\fnm{Luca Maria} \sur{Aiello}}
\author[a,b,+]{\fnm{Giulio} \sur{Cimini}}
\author[c,+]{\fnm{Laura} \sur{Alessandretti}}
\affil[a]{\orgdiv{Physics Department and INFN}, \orgname{University of Rome Tor Vergata}, \orgaddress{\street{Via della Ricerca Scientifica, 1}, \postcode{00133}, \state{Rome}, \country{Italy}}}
\affil[b]{\orgname{Centro Ricerche Enrico Fermi}, \orgaddress{\street{Via Panisperna, 89a}, \postcode{00184}, \state{Rome}, \country{Italy}}}
\affil[c]{\orgdiv{Department of Applied Mathematics and Computer Science}, \orgname{Technical University of Denmark}, \orgaddress{\street{Richard Petersens Plads}, \postcode{2800}, \state{Copenhagen}, \country{Denmark}}}
\affil[d]{\orgdiv{Computer Science Department}, \orgname{IT University of Copenhagen}, \orgaddress{\street{Rued Langgaards Vej 7}, \postcode{2300}, \state{Copenhagen}, \country{Denmark}}}
\affil[e]{\orgname{Pioneer Centre for AI}, \orgaddress{\street{Øster Voldgade 3}, \postcode{2100}, \state{Copenhagen}, \country{Denmark}}}
\affil[$\ast$]{To whom correspondence should be addressed: \href{mailto:antde@dtu.dk}{antde@dtu.dk}}
\affil[$+$]{These authors jointly supervised the work}

\abstract{
In early 2021, the stock prices of GameStop, AMC, Nokia and BlackBerry experienced dramatic increases, triggered by short squeeze operations that have been largely attributed to Reddit's retail investors.
Here we shed light on the extent and timing of Reddit users’ influence on the GameStop short squeeze. 
Using statistical analysis tools with high temporal resolution, we find that increasing Reddit discussions anticipated high trading volumes.
This effect emerged abruptly a few weeks before the event but waned once the community gained widespread visibility through Twitter.
Meanwhile, the collective investment of the community, quantified through posts of individual positions, closely mirrored the market capitalization of the stock.
This evidence suggests a coordinated action of users in developing a shared financial strategy through social media -- targeting GameStop first and other stocks afterwards.
Overall, our results provide novel insights into the role of Reddit users in the dynamics of the GameStop short squeeze.
}
\maketitle
\section{Introduction}

From swinging elections to fueling social movements, instantaneous global communication enabled by online social media can shape leading global phenomena \cite{heidemann_online_2012,aral_introduction_2013,halu_connect_2013,wolfsfeld_social_2013,del_vicario_mapping_2017}.
Financial markets represent a key area where the activity of social media users can have destabilizing effects with worldwide resonance \cite{bouchaud_how_2009}.
Indeed, there is ample evidence that user activity on mainstream platforms like Twitter \cite{bollen_twitter_2011,oliveira_impact_2017} and Google Search \cite{moat_quantifying_2013} not only mirror the mood of the market but can at times anticipate the evolution of stock prices \cite{bollen_twitter_2011,bordino_web_2012,moat_quantifying_2013,curme_quantifying_2014,souma_enhanced_2019,SOUZA2019122343}.
This goes along with the increasingly central role played by retail investors, or non-professional investors, in a sector that was once dominated by institutional funds and large corporations \cite{aramonte_rising_2021,de_silva_losing_2022,van_der_beck_equity_2021}. 

\new{
The dissemination of knowledge on social media stems directly from the dynamic interactions of users \cite{gonzalez-bailon_dynamics_2011}, who actively share knowledge and opinions on financial markets\cite{lillo_how_2015}.
Thus, this medium has the potential to enhance the discovery of information that underlies price formation in financial markets \cite{goldstein_information_2023}. 
Social communication, indeed, supports the price-discovery process \cite{https://doi.org/10.1111/jfir.12310} and improves market efficiency, when information is exogenous \cite{doi:10.1287/mnsc.1120.1678}. 
However, more than rational agents, social media users tend to resemble noisy traders \cite{tetlock_giving_2007}, suggesting that standard models of efficient markets should account for behavioral factors, such as investors' sentiment, public mood \cite{BUKOVINA201618}, cultural traits, psychological biases, social network structure, and information asymmetries \cite{akcay_social_2021} as influential for price formation.
What emerges from this picture is the complex and evolving interplay between social media signals and stock market trends, which enables narratives, ideas, and investment strategies to rapidly spread to a wide audience \cite{https://doi.org/10.1111/jofi.12822,mancini_self-induced_2022}.
Notably, this relationship mainly emerges around specific events, for example when surges in online interactions catalyze trading activity or, conversely, when large price variations spark online conversation \cite{ranco_effects_2015}. 
As such, social media signals tend to capture well transient market trends and stock-specific swings \cite{vassallo_tale_2022}.
}

In early 2021, financial markets were shaken by an unprecedented event: the stock of video-game retail company GameStop (GME) experienced a ``short squeeze'', with a price surge of almost 1625\% within a week \cite{duterme_bloomberg_2023}. 
This financial operation was attributed to activity from users of the social media platform Reddit, particularly the subreddit WallStreetBets (WSB), and was rapidly followed by similar market rallies for other stocks: BlackBerry (BB), AMC Entertainment Holdings (AMC) and Nokia (NOK).
The GME short squeeze suggested that retail investors, by sharing their investment strategies on social media, could potentially challenge the influence of large institutional funds \cite{duterme_bloomberg_2023}.
The Reddit platform could have supported coordination by fostering community-driven discussions \cite{lucchini_reddit_2022,mancini_self-induced_2022,olson_navigating_2015,centola_experimental_2018} focused on shared interests and collective goals. 
On the contrary, mainstream social media like Twitter are primarily used as news feeds \cite{pewresearch_behaviors_2021}, and facilitate coordination only to a limited extent through the use of hashtags \cite{wolfsfeld_social_2013,borge-holthoefer_dynamics_2016}.

The GME short squeeze has been extensively analyzed in the scientific literature. 
Several studies reported a strong correlation between the activity on WSB and the evolution of the GME price throughout the event \cite{mancini_self-induced_2022,lucchini_reddit_2022,sornette_non-normal_2023,lyocsa_yolo_2022,long_i_2023,suchanek_social_2024}. 
However, two major questions remain unanswered regarding the role of WSB users. 
Firstly, it is still unclear if -- and from which specific moment -- the activity of WSB users anticipated the actual GME price surge. 
Secondly, it is unclear if WSB served merely as a platform for financial discourse or as a hub to coordinate a common investment strategy.
In this work, we study the temporal relation among three key signals in the period around the GME event: the stock market movements, the activity on WSB and the broader public attention to the stocks. 
We employ different techniques, from standard Granger Analysis and Multivariate Vector Autoregressive Models to study linear effects, to Detrended Cross-Correlation Analysis (DCCA) and Convergent Cross-Mapping (CCM) to capture the non-linear and complex character of the phenomenon.
Our analysis provides evidence in support of the hypothesis that activity on the WSB community forecast the price surge of GME, BB, AMC and NOK.
Further, we shed light on the temporal dynamics of the GME short squeeze, characterizing its evolution through three distinct phases of online behavior: \textit{Discussion}, \textit{Action}, and \textit{Visibility}.
\section{Results}

\subsection{Reddit activity on GME mirrors changes of GME trading volume.}

We characterise the discourse during the GameStop saga using the volume of GME-related activity on Reddit and on Twitter (see Supplementary Materials, Section 1), the former reflecting the conversation within the community of WSB users and the latter capturing the broader public attention to the stock. 
Following recent literature \cite{curme_quantifying_2014,mancini_self-induced_2022}, we quantify these two signals by measuring the hourly occurrence of the GME ticker in the text of posts and comments on Reddit and Twitter, respectively (see Methods for more details).
The GameStop saga is described by the series of events reported in Table~\ref{table}. 
Interest in GME within the WSB community began to surge as early as December 2020 \cite{lucchini_reddit_2022,mancini_self-induced_2022}. 
However, despite this growing public discourse, the hourly trading volume and price of GME did not exhibit a significant upward trend until approximately 15 days before the short squeeze date of January 27th -- which we associated with Elon Musk's tweet. 
As shown in Figure~\ref{fig1}\textbf{a}, this trend closely resembles the increase in GME-related conversation volume on social media.

This observation suggests a deeper connection between the two signals. 
To shed light on this point, we employ detrended cross-correlation analysis (DCCA). 
This method allows investigating cross-correlations between time series in the presence of non-stationarity \cite{podobnik_cross-correlations_2009}, hence addressing prevalent effects observed in financial time series such as non-linearity, long-term underlying trends and long-range dependencies \cite{mantegna_introduction_1999}.
Specifically, through DCCA, we gauge if fluctuations in GME-related activity are correlated with fluctuations of GME market volume across multiple time scales \cite{curme_quantifying_2014} (see Methods for further details).
We apply DCCA to time series of absolute values of hourly returns for GME occurrences, stock price and trading volume \cite{podobnik_cross-correlations_2009} from December 1st, 2020 to July 1st, 2021, computing the detrended covariances for each combination of signal pairs (see Methods). 
Results shown in Figure~\ref{fig1}\textbf{b} indicate that such detrended covariances follow power-law scaling, pointing to long-term coupling between the signals (see Supplementary Materials, Section 2).
Remarkably, the GME trading volume displays larger detrended covariance with Reddit activity than with GME price, with a cross-correlation coefficient three times larger than the one observed for the other pairs of signals (see Inset Figure~\ref{fig1}\textbf{b}).
This suggests that changes in trading volume are more tied to Reddit discussions, rather than to price movements. 

We further employ Convergent Cross-Mapping (CCM) to corroborate the dynamic interdependence between Reddit activity and trading volume \cite{sugihara_detecting_2012}.
This method quantifies the non-linear coupling between variables by assessing whether past states of one variable $x$ can predict future states of another variable $y$, without assuming model-specific dependencies.
Results of this approach confirm a strong dynamic coupling between Reddit activity and trading volume (see Supplementary Materials, Section 3).
Overall, we can conclude that large variations in Reddit discussions are strongly related to changes in trading volume, as more discussions correspond to a general increase in buying activity and vice versa.

\subsection{The collective position of WSB users on GME amounts to at least 1\% of the stock's market capitalization.}

\begin{figure*}[!ht]
    \centering\includegraphics[width=\textwidth]{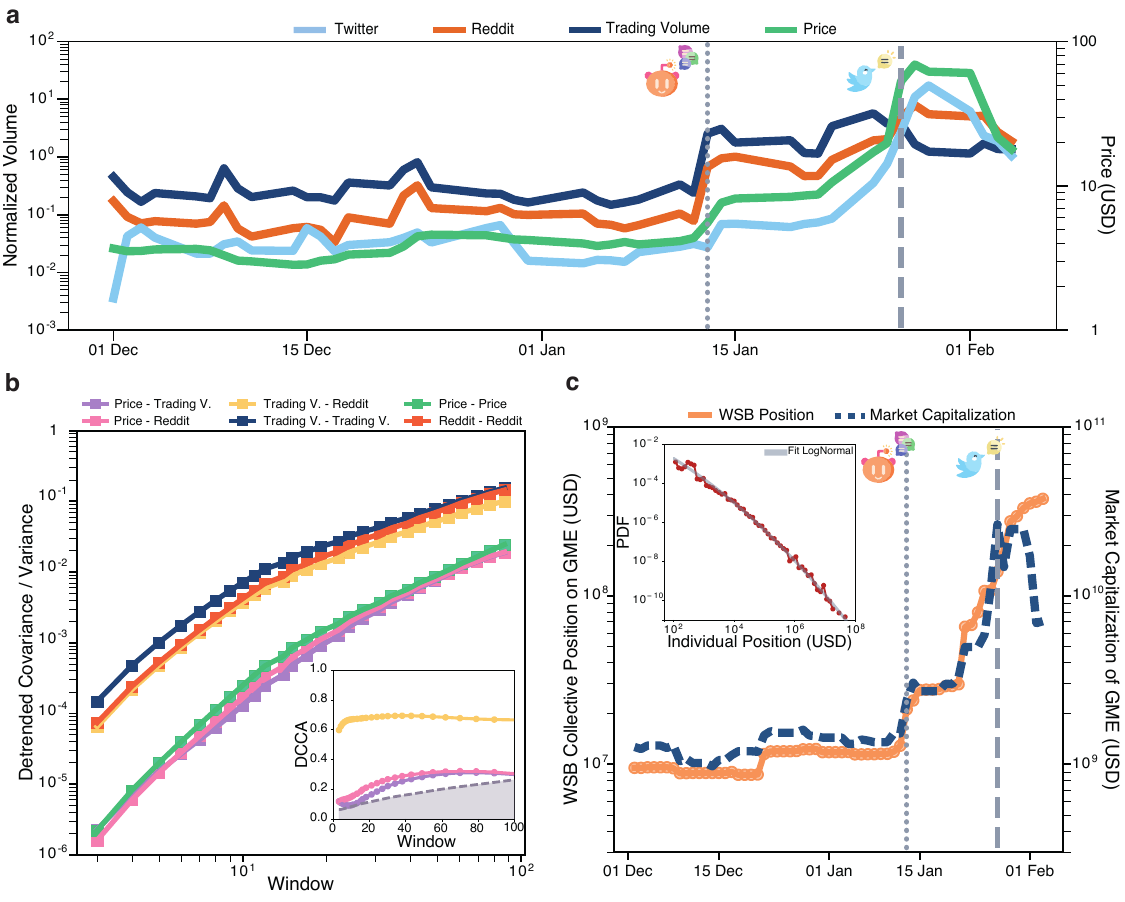}
    \caption{\textbf{The correlation between Reddit activity and GameStop market indicators.}
    \textbf{a)} Daily occurrences of the ``GME'' ticker in conversations on WSB (orange, y-right axis) and Twitter (lightblue, y-right axis), along with the daily trading volume of the GME stock (blue, right y-axis) and its daily closing price (green, left y-axis). 
    The vertical gray dotted line marks January 13th, identified as the WSB-led action's start (see subsequent sections), while the dashed line marks January 27th, when Elon Musk's Tweet broadly publicized the action. 
    To improve the chart's clarity, we applied a 5-day moving average to each signal and normalized the trading volume, Reddit and Twitter signals by their mean (we remark these modifications are not employed in the analysis).
    \textbf{b)} Values of detrended variance for price (green), trading volume (blue) and Reddit activity (red) signals, and values of detrended covariance for the pairs price / trading volume (purple), price / Reddit activity (pink), trading volume / Reddit activity (yellow), as a function of the time window considered.
    Inset: Detrended Cross Correlation Coefficients for paired combinations of GME signals (same color schemes of the main panel). 
    The shaded gray area represents the 95\% confidence interval for correlation values obtained from pairs of independent signals.   
    \textbf{c)} Daily value of the WSB posted collective position on GME (orange, y-right axis) and GME Market Capitalization (blue, y-left axis). 
    Inset: Distribution of values of users' individual position on GME (red) alongside the Log-Normal fit (grey line, $\mu = 4.96$, $\sigma = 3.63$).
    }
    \label{fig1}
\end{figure*}

In WSB, some users provide evidence of their financial investments by posting screenshots of their accounts taken from trading platforms, which attest their positions and trading activity \cite{lucchini_reddit_2022}.
We analyse such screenshots to capture the actual engagement of WSB users with the GME stock \cite{curme_quantifying_2014,mancini_self-induced_2022}, using computer vision techniques (see Methods for more details and Supplementary Materials, Section 4).
We observe that the values of the individual positions on GME follow a Log-normal distribution (see the Inset of Figure~\ref{fig1}\textbf{c}). 
By summing the positions across all users, we obtain the posted collective investment, which provides a lower-bound estimate of the community stake in GME, as it includes only the retail investors who have publicly shared their positions.
We find that this posted collective position grew significantly 15 days prior to the squeeze. 
This finding complements the observation that the number of WSB users posting screenshots grew in the period preceding the squeeze \cite{lucchini_reddit_2022}, likely indicating that we have more users investing in GME.
Notably, the growth of the posted collective position closely follows the market capitalization of GME (see Figure~\ref{fig1}\textbf{c}).
We estimate that the posted collective investments of Reddit users represented -- on average over the time period considered -- at least the 1\% of the total market capitalization of GME (see Figure~\ref{fig1}\textbf{c}).

\subsection{Granger analysis identifies three distinct phases of online behavior.}

We now test the hypothesis that changes in GME-related activity on social media \textit{anticipated} changes in GME trading volume around the short squeeze. 
Specifically, we employ the Granger test to assess the chronological anticipation $x\to y$: whether a signal $x$ has predictive power on another signal $y$, building a linear regression model for $y(t)$ leveraging past information on both $y$ and $x$ from $t-\Delta t$ up to $t-1$.
The Granger Index \cite{geweke1984measures} quantifies the test result as the ratio of predictive accuracy between the model and a baseline that excludes past information derived from $x$ (see Methods for more details).
As $x$ and $y$ we consider the possible pairs of signals among GME trading volume and GME-related discussions on Reddit and Twitter. 
We consider hourly values during market opening times, within a 15-day time window (corresponding to 120 points per time series -- see Methods) between January 5 and February 5.
This choice of window size ensures that we capture short-term patterns that are relevant to our case study \cite{vassallo_tale_2022} and at the same time that we have enough statistics to draw robust conclusions from the Granger analyses, even in regimes of small coupling \cite{ramos_minimum_2017}. Results are, however, robust to different choices of window length (see Supplementary Materials, Section 5).
After testing the stationarity of the time series within the specified windows (see Supplementary Materials, Section 5), we apply a daily moving average (the results remain robust when we do not apply the moving average, see Supplementary Materials, Section 5) and compute the hourly log returns of the signals.

\begin{figure*}[!ht]
    \centering\includegraphics[width=\textwidth]{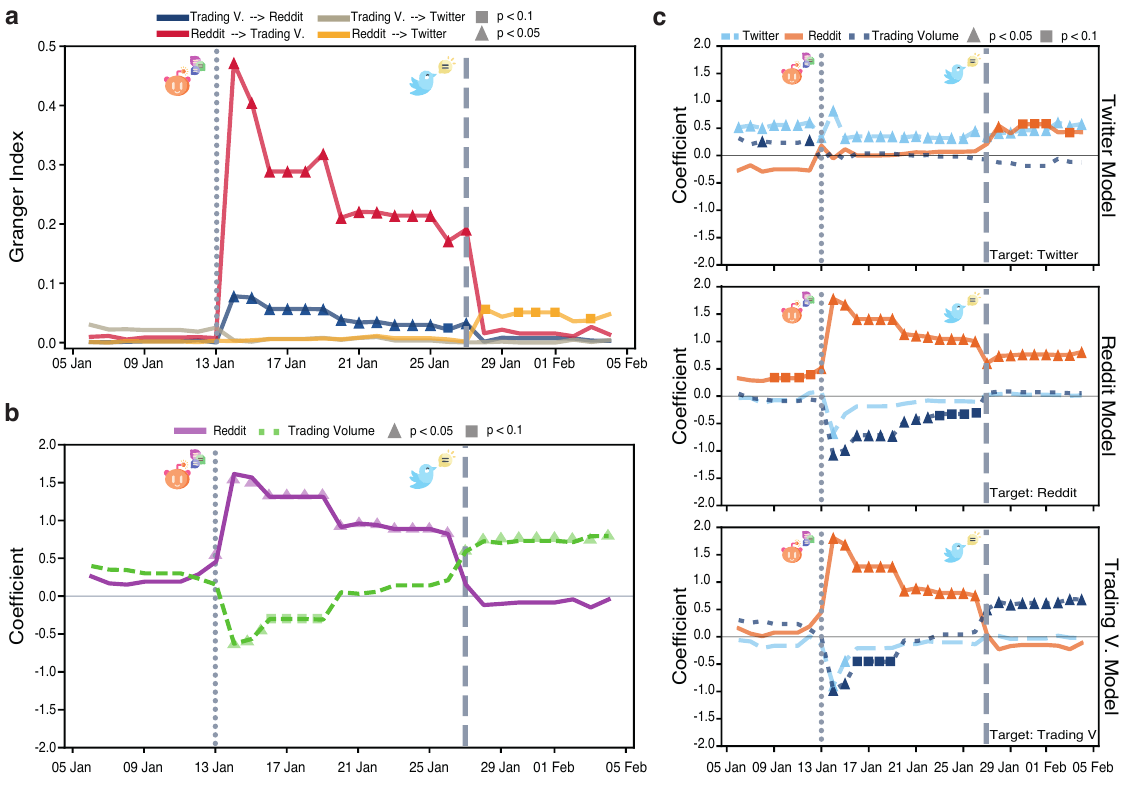}
    \caption{\textbf{Reddit activity anticipates trading volume during the short squeeze.}
    In all the following subplots the vertical gray dotted line indicates the beginning of the \textit{Action} phase (13 January 2021), whereas the vertical gray dashed line corresponds to the Tweet by Elon Musk (27 January 2021) that brought the squeeze to the public attention. \new{Triangles correspond to p-values $<0.05$, squares to p-values $<0.1$.}
    Each point is computed considering time series spanning the 15 preceding days.
    \textbf{a)} Granger Index capturing the predictive power of a signal on another (with a lag of 1 hour) for the following pairs: Trading Volume-to-Reddit (blue), Reddit-to-Trading Volume (red), Trading Volume-to-Twitter (gray) and Reddit-to-Twitter (yellow).
    \textbf{b)} Coefficients of the Granger Model predicting Trading Volume, capturing the size effects of Reddit (magenta) and Trading Volume (green) activities.
    \textbf{c)} Each panel shows the coefficients of a Multivariate Vector Autoregressive Model predicting Twitter (top), Reddit (middle), and Trading Volume (bottom) activities, corresponding to antecedent values of Reddit (solid orange), Twitter (dashed lightblue), and trading volume (dotted blue).
    }
    \label{fig2}
\end{figure*}

Figure~\ref{fig2}\textbf{a} shows the resulting Granger Index for $\Delta t=1$, meaning that the model uses $x(t-1)$ to predict $y(t)$. 
In the plot, triangles correspond to statistical significant values (p-values $<0.05$).
Size effects are instead captured by the coefficient of the regression of the independent variable, reported in Figure~\ref{fig2}\textbf{b} for the Reddit-to-Trading Volume direction (see Supplementary Materials, Section 5 for other values). 
Results highlight three distinct phases of online behavior (see Figure~\ref{fig2}\textbf{a}).

\paragraph{Phase 1.} Before January 13th, the Granger Index and size effects are not significantly different from zero (p-values $>0.1$) for all combinations of signals, meaning that no anticipatory effects are observed.  

\paragraph{Phase 2.} On January 13, the Granger Index capturing the anticipatory power of the Reddit activity on trading volume has a sharp transition to a very high value of 0.47 (p-values $<0.05$). 
At the same time, size effects grow up to 1.5, meaning that social media activity has a large positive effect in the prediction of market trends.
Simultaneously, the predictive power of the trading volume on Reddit \new{is} significant; however, the Granger Index in the Reddit-to-Trading Volume direction is roughly three times larger than in the opposite Trading Volume-to-Reddit direction.
These indices remain significant -- although steadily declining -- for about 12 days. 
\new{
The p-values are reported in Supplementary Materials, Section 5.
}

%
\paragraph{Phase 3.} On January 27th, the Tweet ``Gamestonk!!'' by Elon Musk brought the GME short squeeze to the public attention. 
On this date, the anticipatory power of Reddit activity on the trading volume (and viceversa) \new{is} non significant, and size effects vanish. 
Instead, Reddit activity starts to anticipate the activity on Twitter: once they became informed about the short squeeze, Twitter users turned to Reddit for information about GME.

\bigskip
The three phases identified above suggest the following sequence of events. Prior to January 13th, WSB users were discussing about GME but did not play a significant role in the market. 
Afterwards, from January 13th to 27th, their coordinated action of buying and holding GME shares began to influence the market. 
Finally, after January 27th, the operation gained public attention, the market was influenced by a much broader audience and other market players -- eclipsing the role of WSB users. At the same time, a larger horde of people joined the WSB community, fundamentally altering the nature of discussions \cite{lucchini_reddit_2022,mancini_self-induced_2022}. 
To reflect their distinct nature, in the following, we refer to these three phases as the \textit{Discussion}, \textit{Action}, and \textit{Visibility} phases, respectively.

\subsection*{Reinforcing effects among Reddit, Twitter and Trading Volume.}

The analysis in the previous section considers only pairwise relations among signals.
In order to capture how the mutual relationship between all three signals under study change over time, we employ a Multivariate Vector Autoregressive model.
The method generalises the Granger analysis to multiple signals: it entails fitting a linear model that predicts a target output signal $y$ at time $t$, using past information from $t-1$ to $t-\Delta t$ on $y$ itself plus one or more input feature signals $\{x_1, x_2, x_3, \dots\}$. 
The coefficients of the model capture how the individual input signals contribute to predicting future values of $y$ (see Methods for more details). 
In particular, a positive coefficient for $x_i$ indicates a positive reinforcing relationship of $x_i$ on $y$, while a negative coefficient an opposite and counteracting effect.
We apply the method to study the hourly log-returns of Twitter activity, Reddit activity, and trading volume. As we did for the Granger analysys, we fitted the model within each 15-day period ending on a day between January 5th and February 5th, for $\Delta t = 1$ (results for other window period lengths are reported in the Supplementary Materials, Section 5).

Figure~\ref{fig2}\textbf{c} shows the coefficients that capture the predictive power of the three signals on the future values of Twitter (top plot), Reddit (mid plot), and trading volume (bottom plot).
First, we observe a self-reinforcing dynamic within Reddit and Twitter throughout the whole period under study, where increased activity within the social network leads to further increases.
This is evident from the consistently positive and significant coefficients of the Reddit-to-Reddit and Twitter-to-Twitter relation.
During the \textit{Action} phase, the self-reinforcing nature of Reddit is even more pronounced. 

Secondly, during the \textit{Action} phase, the coefficient capturing the Reddit-to-Trading Volume relation \new{is} significant and positive, implying that the increase of activity on Reddit is followed -- within the next hour -- by an increase in trading volume. 
This is coherent with the size effects observed in the Granger analysis.
Notably, in this phase the magnitude of the Reddit-to-Trading Volume coefficient is similar to the magnitude of the Reddit-to-Reddit coefficient (the two can be directly compared since we are considering the log-returns of the signals).

Finally, during the \textit{Visibility} phase, the trading volume starts following a self-reinforcing dynamics (as indicated by the significant and positive coefficient for the Trading Volume-to-Trading Volume relation), while the Reddit-to-Trading Volume coefficient \new{is} non significant.
In this phase, the Reddit-to-Twitter coefficient aligns in both magnitude and direction with the Twitter-to-Twitter coefficient, again highlighting that public attention was directed towards WSB.

By conducting this analysis at various lags $\Delta t$, we found that the coefficients at $\Delta t=1$ hour are larger than the others (see Supplementary Materials, Section 5), indicating the greater significance of short time scales in the entwined dynamics of social media and markets.
Interestingly, the typical reply speed, defined as the average time interval between a comment and its direct responses, varies between $\sim2.5$ hours during the \textit{Discussion} phase to $\sim0.5$ hours during the \textit{Visibility} phase (see Supplementary Materials, Section 6). 
Hence, the influence of Reddit on the market occurs on a timescale closely aligned with that of the unfolding conversations.

%
\subsection*{The role of WSB users on other short squeezes.}

Alongside GME, also BB, AMC, and NOK experienced a short squeeze in January-February 2021.
In this section, we investigate the interplay between discussions on WBS and these financial events, employing the techniques introduced in the previous sections. 
Our analysis focuses on the occurrences of the BB, AMC and NOK tickers on WSB posts and comments, against the trading volumes of the respective stocks.

\begin{figure*}[!ht]
\centering\includegraphics[width=1.0\textwidth]{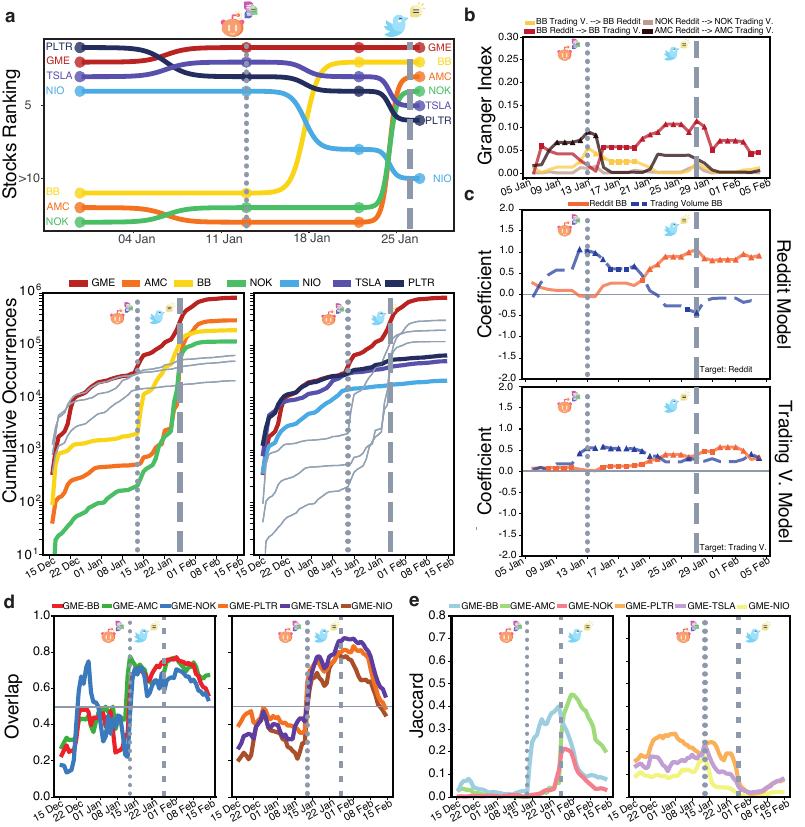}
    \caption{\textbf{BB, AMC, NOK and the role of GME-engaged users.}
    In all the following subplots the vertical gray dotted line indicates the beginning of the GME \textit{Action} (13 January 2021), whereas the vertical gray dashed line corresponds to the Tweet by Elon Musk (27 January 2021). \new{Triangles correspond to p-values $<0.05$, squares to p-values $<0.1$.}
    \textbf{a)} (top) Rank chart of stock tickers on WSB based on their average daily occurrence within each 15-days window.
    \textbf{a)} (bottom) Cumulative number of occurrences of a given stock on WSB.  
    \textbf{b)} Granger Index for possible combinations of BB, NOK, AMC signals.
    \textbf{c)} Coefficients of the Multivariate Vector Autoregressive Model for BB-related activity on Reddit (top panel) and BB trading volume (bottom panel), related to the Reddit (solid orange) and trading volume (dashed blue) signals.
    \textbf{d)} Overlap index for pairs of stocks, computed over a 5-day window with a one-day shift, measuring the proportion of users discussing each pair of stocks. We observe in both panels an increasing overlap after the starting of the GME rally.
    \textbf{e)} Jaccard index for pairs of stocks, computed over a 5-day window with a one-day shift, measuring the proportion of users discussing both pair of stocks.
    }
    \label{fig3}
\end{figure*}

By the end of 2020, BB, AMC and NOK ranked below position 10 among the stocks that were most discussed on WSB, while Palantir (PLTR), Tesla (TSLA) and NIO Inc. (NIO) occupied the top ranks together with GME (see Figure~\ref{fig3}\textbf{a}, top panel).
Interestingly, during the GME \textit{Action} phase, we observe a fast ramp up of BB, AMC and NOK which -- by mid January 2021 -- became the most discussed stocks, just below GME.

The Granger analysis reveals that, during the GME \textit{Action} phase, the Reddit-to-Trading Volume anticipatory relation is only significant for BB (see Figure~\ref{fig3}\textbf{b}).
Interestingly, the Granger Index for BB \new{is} statistically significant only five days after it does so for GME. 
The analysis for AMC and NOK, instead, does not lead to significant conclusions (see Supplementary Materials, Section 5), possibly due to the less attention they received on Reddit (see Figure~\ref{fig3}\textbf{a}).
We refer the reader to the Supplementary Materials, Section 2 for the analyses of long-range detrended cross-correlations (DCCA) and Convergent Cross-Mapping (Supplementary Materials, Section 3) between trading volume and the frequency of occurrences for BB, AMC and NOK.

We now consider the Multivariate Autoregression analysis of BB (see Supplementary Materials, Section 5 for the analysis of NOK and AMC). 
As Figure~\ref{fig3}\textbf{c} shows, we find that the Reddit-to-Trading Volume anticipatory effect is reinforcing, as demonstrated by a significant and positive model coefficient. 
We observe, however, a different behaviour compared to the GME case. 
At the beginning of the \textit{Action} phase, the Trading Volume has a positive reinforcing effect on both Reddit and itself; at the same time, Trading Volume is much better predicted by itself than by Reddit. 
This trend reverses one week later, when Trading Volume loses all its explanatory power, while Reddit enters into a positive relationship with itself and Trading Volume.

But how do BB, AMC and NOK gained popularity among WSB users ?
To study this phenomenon, we compute the overlap coefficient (size of the intersection over size of the smaller set) and Jaccard index (size of the intersection over size of the union) among the set of users who talk about GME and those who talk about other stocks (see Figure~\ref{fig3}\textbf{d} and \ref{fig3}\textbf{e}).
On January 13th, concurrently with the beginning of the \textit{Action} phase, we observe a sharp increase in the overlap between users discussing GME and other stocks. 
After January 13th, the overlap index has a plateau at a high value close to $0.8$, signaling an expansion of discussions among WSB users. 
Initially focused on GME, they broadened their conversations to include various other stocks.
By studying the Jaccard index, we notice that up to 40\% of users interested in GME focused initially also on BB; then, as the short squeeze approached, they turned to NOK and AMC.
During the same period, the number of users endorsing stocks that were top ranked before December (PLTR, TSLA, NIO) decreased. 
When we compute these indices, the group of users who talk about the other stock (not GME) is always smaller, while the larger group is mostly made up of users discussing GME (see Supplementary Materials, Section 7).
Additional analyses based on the raw number of users discussing the various stocks show that, prior to the start of the GME rally, users stopped discussing NIO, and during the rally started focusing on BB, NOK and AMC (see Supplementary Materials, Section 7). 
Simultaneously, mega-threads dedicated to GME appear on WSB \cite{pei_searching_2014,mancini_self-induced_2022}, providing additional visibility to other stocks that GME users were also talking about.

These findings suggest that WSB users engaged with GME also endorsed BB, AMC and NOK. 
Approaching the GME short squeeze, and concurrently with the widespread recognition of WSB on a global level, these stocks emerged as the most popular choices and ended up as short squeeze targets.

\section{Discussion}
In January 2021, the activity of WSB users anticipated financial trends.
The high temporal resolution of our analysis enabled us to identify three distinct phases of online behavior and engagement with the stock market, characterized by \textit{Discussion}, \textit{Action} and \textit{Visibility}, respectively.
First, we observed that the strong anticipatory relation between WSB users activity and changes in GME trading volume emerged sharply on January 13th.
This suggests a sudden transition between a phase characterized by \textit{Discussion} and one characterized by financial \textit{Action}.
Second, we found that -- at around the same time -- the value of the collective position on GME started to closely mirror the evolution of the market capitalization of GME.
We estimated that the size of the total position of WSB users amounted to at least 1\% of GME market capitalization, demonstrating a tangible engagement of the WSB community in the stock market.
This suggests that other market players joined the short squeeze after it was initiated by WSB users, despite having access to information about GME that was presumably not inferior to that of the WSB users \cite{pelaez2021David}.
After the price hike of GME and a Tweet by Elon Musk on January 27th, the short squeeze became widely known, and a broader public started to follow Reddit conversations.

Our study comes with limitations. 
First, our findings refer to a specific case study, the GME short squeeze, which so far represents a unique event due to its wide resonance.
If and when such events will occur again, we could understand whether the patterns we have observed can be generalized to different contexts.
Secondly, our analysis primarily relies on Granger test, a standard statistical framework that relies on linear models.
As such, it may oversimplify complex relationships, such as those involved in collective action between social networks and financial markets, where feedback loops and non-linear dynamics can occur.
Recognizing this limitation, we have explored nonlinear techniques; notably, an analysis based on Convergent Cross-Mapping \cite{sugihara_detecting_2012} corroborates our main results (see Supplementary Materials, Section 3).
Third, due to limitation of the available data, our analysis focused solely on Reddit and Twitter, neglecting offline interactions or happening on other online platforms (\eg, Telegram or Discord), as well as the online/offline actions of institutional investors. 
Finally, the proxy used to estimate the size of WSB collective position on GME, specifically from user screenshots, are inherently partial and noisy.
We also have no information on the identities of the investors behind the screenshots, so we cannot exclude the potential interference by institutional investors in the financial discourse on Reddit.

Despite such limitations, we have provided extensive evidence regarding the possible influence exerted by Reddit users on the GameStop short squeeze.
However, determining a genuine causal link -- that Reddit users directly fueled the price rally -- is not possible with the data we can access. 
Factors such as the true portfolio composition of each Reddit user and the potential psychological contagion within Reddit discussions add uncertainty for the attribution of direct causation.

Overall, our results add further evidence that digital platforms can catalyze collective action and influence real-world outcomes.
In an era characterized by rapid technological advancements, the pace of our daily lives is accelerated to an unprecedented rate \cite{lorenz-spreen_accelerating_2019}. 
This phenomenon has also impacted the financial landscape, where the emergence of commission-free trading platforms started a new age of dynamism and accessibility \cite{duterme_bloomberg_2023}.
Our findings highlight the importance of investigating the dynamic interplay between online engagement and stock market movements at an hourly scale (or even finer) \cite{bordino_web_2012,curme_quantifying_2014,bollen_twitter_2011,moat_quantifying_2013}, as the coordinated action of retail traders through social media has become a no-longer-negligible market player.

\section{Methods}
\subsection{Data Descriptions}
We gathered Reddit conversation data from Pushshift \cite{baumgartner_pushshift_2020}, an API that regularly copies activity data of Reddit. 
We queried the service to retrieve information about posts and comments on WSB from December 01, 2020 to July 1, 2021. 
The dataset was cleaned by removing posts/comments by Reddit bots (see Supplementary Materials, Section 1) as well as by "[deleted]" users, who deleted their account at the time Pushshift API collected data.
To obtain the occurrences of a given stock within WSB conversations, we counted how many times the ticker symbol of the stock (\eg, `NOK' for Nokia) appears as a regular expression in the raw text of a post/comment \cite{mancini_self-induced_2022}.

Within the WSB community, a common practice consists in posting financial investments by sharing screenshots of open positions, primarily gains, losses, or orders.
These ``committed'' screenshots were identified and classified using computer vision techniques by Lucchini \ea \cite{lucchini_reddit_2022}.
Using Tesseract, an open-source optical character recognition \cite{tesseract}, we extract textual information from these screenshots, converting image-based text into editable and searchable content.
We focused on screenshots containing keywords such as `gme,' `gamestop,' `investing,' and `value,' identified through manual inspection, and subsequently narrowed down the sample to 6525 screenshots from November 1, 2020 to February 04, 2021.
After parsing the structures, we isolated dollar-sign-prefixed numbers, determining the final value of the position as the highest extracted value.
However, such value may represent trading volume or the price of GME taken from online trading applications, rather than an actually investment.
To address this, we filtered out screenshots with values below $10^2$ and above $10^6$, as these are likely indicative of price or trading volume. We manually examined values above $10^6$ to validate our findings.
\new{
Additionally, we manually validated 1000 screenshots using log-transformed data, computing the logarithms of the ratios between true and estimated values. We determined that our procedure extracts the correct value with an accuracy of 0.85, where accuracy is based on a binary classifier where correctness is determined by applying a threshold of 0.05 to the absolute logarithmic ratios (see Supplementary Materials, Section 4).
The root mean square error (RMSE) of the logarithmic ratios between true and estimated values is 0.47, and the point estimate (i.e., the average ratio for a single screenshot) amounts to -0.07. 
This indicates that the automated predictions slightly overestimate the actual values (on average +17\%). 
However, this difference is more pronounced at screenshot values below $10^3$, with a relative logarithmic error of -0.1, while it approaches zero at higher values. 
Overall the automated extraction procedure leads to a total discrepancy of approximately +0.2\% in the WSB posted collective position.
}
Note that this validation was intended solely to assess the error of the employed framework in estimating the collective financial position -- the incorrect estimates were not replaced. 
Indeed, missing posts, and possibly fake ones, represent a much larger source of error, which we cannot control.

We measured Twitter interest towards GME using Twitter API for academic research.
From this service, we collected tweets with the \#GME or \#Gamestop hashtags posted between December 1, 2020 and February 5, 2021 \cite{lucchini_reddit_2022}.
Data on stock close price and traded volumes were retrieved from the API service of \url{https://polygon.io}.
Data on stock close price and traded volumes were retrieved from the API service of \url{https://polygon.io}, wherein the price data reflects stock split adjustments that occurred during the stock split in 2022.
All signals were taken into account exclusively during market hours (9 a.m. to 5 p.m.). 
When analyzing the changes of a particular signal, denoted as $x(t)$, we compute its hourly logarithmic return as $r(t) = \ln[x(t+1)/x(t)]$.
\subsection{Detrended Cross-Correlation Analysis}

Detrended Cross-Correlation Analysis is a statistical technique designed to explore long-range dependencies between two time series, by studying how the fluctuations of local trends evolve across temporal scales \cite{podobnik_statistical_2011,podobnik_cross-correlations_2009}.
The underlying idea is to decompose two signals into smaller overlapping segments using a given time window, compute the covariance between the two detrended signals within each segment, average the results obtained over the various segments, and finally repeat all the steps for windows of different lengths to assess potential scale relations.
Following Podobnik \ea\cite{podobnik_statistical_2011},
we consider two time series $x(\tau)$ and $y(\tau)$ of equal length $T$ and compute the integrated signals $I^{(x)}(t) = \sum_{\tau = 1}^{t}x(\tau)$ of each time series.
For the analysis at window length t, we divide the signals into $T-t$ consecutive segments (window $t$), each containing $t+1$ values.
In each segment that starts at $i$ and ends at $i+t$, we define the local trend as the ordinate of a linear least-squares fit $S_{i}^{(x)}(t)$ and remove it from the integrated windowed signal $I_{i}^{(x)}(t)$ to compute the residuals. 
We define the covariance of the residuals in each segment as
\begin{equation}
f_{DCOV} (i,t) =\frac{1}{t-1}\sum_{\tau = i}^{t+i} (I^{(x)}(\tau)-S_{i}^{(x)}(\tau))(I^{(y)}(\tau)-S_{i}^{(y)}(\tau))\,,
\end{equation}
and if $x = y$ we have the variance of the residuals $f^2_{DVAR} (i,t)$.
Then, the detrended covariance $F^{(x,y)}_{DCOV}(t)$ is defined as the average covariance of the residuals over the overlapping segments at the given window $t$: $F^{(x,y)}_{DCOV}(t) = \frac{1}{T-t}\sum_{j = 1}^{T-t}f_{DCOV} (j,t)$. 
In our case, the time series have temporal length $T \sim 10^3 $ and the windows considered are logarithmically spaced, each containing 30 points.
The detrended cross-correlations coefficient is finally defined as
\begin{equation}
F_{DCCA}(t) = \frac{F^{(x,y)}_{DCOV}(t)}{\sqrt{F^{(x,x)}_{DVAR}(t)}\sqrt{F^{(y,y)}_{DVAR}(t)}}\,.
\end{equation}
If two series are power-law cross-correlated, then the detrended covariance as a function of the window length $t$ follows a power law, $F^{(x,y)}_{DCOV}(t)\sim t^\lambda $ and the magnitude of $F_{DCCA}$ indicates the correlation strength \cite{podobnik_statistical_2011}.
The exponent $\lambda$ identifies whether this cross-correlation is short-range ($\lambda = 0.5$) or long-range ($\lambda > 0.5$).
If two time series are short-range correlated or cross-correlations is absent, $F_{DCCA}$ could be non zero and slightly be dependent on window $t$ due to size effects \cite{podobnik_statistical_2011}.
To identify statistically significant values, we compare the obtained cross-correlations against those obtained for two independent random signals (see Inset Figure \ref{fig1}\textbf{c}, gray area)\cite{podobnik_statistical_2011}.
In our case, we observe values of cross-correlation always exceeding the confidence interval, and scaling exponents $\lambda > 0.5$, pointing to power-law cross-correlations (see Supplementary Materials, Section 2).
However, it is crucial to interpret the fitting of the exponents as an exercise due to the limitation of short time series.
We further corroborate our findings by estimating the Hurst exponent through R/S analysis (see Supplementary Materials, Section 2).
\subsection{Granger test and Multivariate Vector Autoregressive Model}

The Granger test is commonly used in time series analysis to determine whether the predictability of $y(t)$ in a given model decreases when the time series $x(t)$ is excluded from that model \cite{bordino_web_2012}. 
The key idea is that when both past values of $x$ and $y$ are incorporated rather than just past values of $y$, the prediction of $y$ improves, thus indicating that $x$ Granger-causes $y$.
In our case, we consider a vector autoregressive model of order $\Delta t$
\begin{align}
y_t &=  \widetilde{a_0}+\sum_{l=1}^{\Delta t}\widetilde{a_l}y_{t-l}+\widetilde{\epsilon_t} \qquad \text{and} \nonumber \\
y_t &=  \widehat{a_0}+\sum_{l=1}^{\Delta t}\widehat{a_l}y_{t-l}+\sum_{l=1}^{\Delta t}\widehat{b_l}x_{t-l}+\widehat{\epsilon_t}\,,
\end{align}
where the first equation is the restricted model and the second the full model.
After fitting these models, the p-values are computed from the F-statistic, which reflects the difference in the residual sum of squares ($RSS$). 
These p-values indicate the likelihood of obtaining results as extreme as the observed F-statistic under the assumption that the restricted model is correct.
Size effects are given by the values of the regression coefficients $\widehat{a_l}$ and $\widehat{b_l}$, which capture the contribution of the lagged variables $y$ and $x$ in predicting $y$.
The Granger Index, on the other hand, focuses only on the predictive accuracy of the full model in relation to the restricted model, defined as $GI = \ln(RSS_{restricted}/RSS_{full})$.
To conduct the Granger test, we need first to evaluate the stationarity of signals within the specified windows. 
The results of the Augmented Dickey-Fuller test are reported in Supplementary Materials, Section 5.

Following the full model of the Granger test, we consider a Multivariate Vector Autoregressive model to analyze relationships and dependencies among two or more time series simultaneously.
The significance of a particular regression coefficient is established by evaluating its corresponding p-value, computed similarly to the Granger analysis.
Both for Granger and Multivariate Vector Autoregressive Model, we analyzed the data by sliding it within 15-day windows, shifting by one day each time. 
The resulting length of the time series is thus of 120 points (8 hourly points per day) -- a reasonable statistics to perform linear regression analysis \cite{austin_number_2015,box_time_2008,ramos_minimum_2017}.

\section*{Declarations}
\begin{itemize}
\item Data Availability:
Reddit conversation data used in this study (the same used by Mancini \ea \cite{mancini_self-induced_2022} and Lucchini \ea \cite{lucchini_reddit_2022}) can be retrieved from the Pushshift API at \url{https://www.reddit.com/r/pushshift/} and were collected in December 2022.
Twitter data used in this analysis are the same used by Lucchini \ea\cite{lucchini_reddit_2022}, and were retrieved from Twitter API before April 2022.
Stock market data used in this analysis can be retrieved from the Polygon.io API at \url{https://polygon.io} and were collected in December 2022, while the GME Market Capitalization was collected in September 2023.
\item Code Availability:
The data and code to reproduce the analysis is released on \href{https://github.com/RiegelGestr/WSBCausality}{GitHub}.
For inquiries, please contact A.D. \url{antde@dtu.dk}.
\item Acknowledgements
A.D. acknowledges the helpful discussions on Twitter and screenshot data analysis with Lorenzo Lucchini.
A.D. acknowledges the helpful discussions on Reddit conversations analysis with Riccardo Di Clemente, Anna Mancini and Gabriele Di Antonio.
A.D. acknowledges Danish Data Science Academy for their funding support ("Grant ID DDSA-V-2023-008").
L.M.A. acknowledges support from the Carlsberg Foundation through the COCOONS project (CF21-0432).
G.C. acknowledges support from “Deep ’N Rec” Progetto di Ricerca di Ateneo of University of Rome Tor Vergata. 
\item Author Contributions:
A.D. and L.M.A gathered the data.
A.D. performed the analysis and made the figures. 
L.A. and G.C designed the analysis and supervised the project.
All authors discussed the results and contributed to the final manuscript.
\item Competing Interests:
The authors declare no competing interests
\end{itemize}
\bibliography{bibliography.bib}
\begin{table}[h!]
    \centering
    \begin{tabular}{llp{0.5\textwidth}}
    \toprule
    \textbf{Date} & \textbf{Event} & \textbf{Description} \\
    \midrule
    2020-12-08 & GME Q3 Earnings Reports & GameStop's Q3 earnings highlighted a significant 257\% surge in e-commerce revenues. \\
    2021-01-11 & New Board of Directors & GameStop revealed a new Board of Directors, including several experts in e-commerce, signaling a new leadership direction. \\
    2021-01-13 & GME Stock Rise Begins & GME stock price started rising significantly, increasing by around 60\% in a single day. \\
    2021-01-19 & Citron Tweet & Citron Research criticized GameStop's stock valuation and forecasted a price drop, belittling GME buyers on Twitter. \\
    2021-01-22 & BB Stock Surge & BlackBerry stock surged by 30\%. \\
    2021-01-27 & Elon Musk Tweet & Elon Musk tweeted, ``Gamestonk!'' with a link to the subreddit r/WallStreetBets. GME stock surged over 135\%. \\
    2021-01-28 & Robinhood Trading Restrictions & Robinhood and other brokerage platforms restricted the buying of GME, AMC, BB, and NOK stocks, leading to a temporary price crash. AMC stock rose by 300\%, BlackBerry by 32\%, and Nokia stocks surged by 30\%.\\
    \bottomrule
    \end{tabular}
    \caption{Major events relevant to the GameStop (GME) short squeeze and the subreddit r/WallStreetBets (WSB).}
    \label{table}
\end{table}

\pagebreak[2]
\clearpage
\renewcommand{\figurename}{Supplementary Figure}
\renewcommand{\tablename}{Supplementary Table}
\setcounter{figure}{0}
\renewcommand{\thefigure}{S\arabic{figure}}
\section*{Supplementary Materials, Section 1: Dataset Information}\label{sm_dataset_info}
\begin{table}[h!]
	\centering
	\begin{tabular}{lp{0.6\textwidth}}
		\toprule
		{\textbf{Column}} & {\textbf{Description}}\\
		\midrule
		Author &  Username\\
		Author ID & ID that uniquely identifies each Reddit user\\
		Comment ID & ID that uniquely identifies each comment\\
		Submission ID &	ID of the post under which the comment was made\\
		Parent ID &	ID of the post or ID of the comment to which the given comment is a reply\\
		Text & Text of the comment\\
		UTC & Epoch Unix timestamp of the comment\\
		\hline 
	\end{tabular}
    \caption{Metadata downloaded from Pushshift for each Reddit comment.
    } 
\label{tab_sm_data_comments}
\end{table}
\begin{table}[h!]
	\centering
	\begin{tabular}{lp{0.6\textwidth}}
		\toprule
		{\textbf{Column}} & {\textbf{Description}}\\
		\midrule
		Author &  Username\\
		Author ID & ID that uniquely identifies each Reddit user\\
		Submission ID & ID that uniquely identifies each submission\\
		Text &	Text of the submission\\
		Title &	Title of the submission\\
		UTC & Epoch Unix timestamp of the submission\\
		\hline 
	\end{tabular}
    \caption{Metadata downloaded from Pushshift for each Reddit submission.
    }
\label{tab_sm_data_subs}
\end{table}
\begin{table}[h!]
	\centering
	\begin{tabular}{c}
		\toprule
		{\textbf{Bot name}}\\
		\midrule
	    WSBVoteBot\\
        RemindMeBot\\
        Generic\_Reddit\_Bot\\
        ReverseCaptioningBot\\
        LimbRetrieval-Bot\\
        NoGoogleAMPBot\\
        RepostSleuthBot\\
        GetVideoBot\\
        CouldWouldShouldBot\\
		\hline 
	\end{tabular}
    \caption{Reddit bots removed.
    } 
\label{tab_sm_bots}
\end{table}
\begin{table}[h!]
	\centering
	\begin{tabular}{lp{0.45\textwidth}}
		\toprule
		{\textbf{Column}} & {\textbf{Description}}\\
		\midrule
		Time &  Day timestamp of Stock Index\\
		Open & Daily Opening Value of Stock Index\\
		Close & Daily Closing Value of Stock Index\\
		High & Daily High Value of Stock Index\\
		Low & Daily Low Value of Stock Index\\
		Trading Volume & Daily Volume of Stock Transactions\\
		\hline 
	\end{tabular}
    \caption{Metadata downloaded from \url{polygon.io} for each stock ticker.
    } 
\label{tab_sm_stocks}
\end{table}
\begin{table}[h!]
	\centering
	\begin{tabular}{lp{0.45\textwidth}}
		\toprule
		{\textbf{Column}} & {\textbf{Description}}\\
		\midrule
		Time &  Day timestamp of Market Capitalization\\
		Market Capitalization & Daily Market Capitalization\\
		Outstanding Shares & Outstanding shares held by investors\\
		\hline 
	\end{tabular}
    \caption{Metadata downloaded from \url{polygon.io} for GME market capitalization.
    } 
\label{tab_sm_market_cap}
\end{table}
\pagebreak[2]
\clearpage
\section*{Supplementary Materials, Section 2: Detrended Cross-Correlations and Hurst Exponent}\label{sm_hurst}

\begin{figure}[h!]
    \centering
    \includegraphics[width=1.0\textwidth]{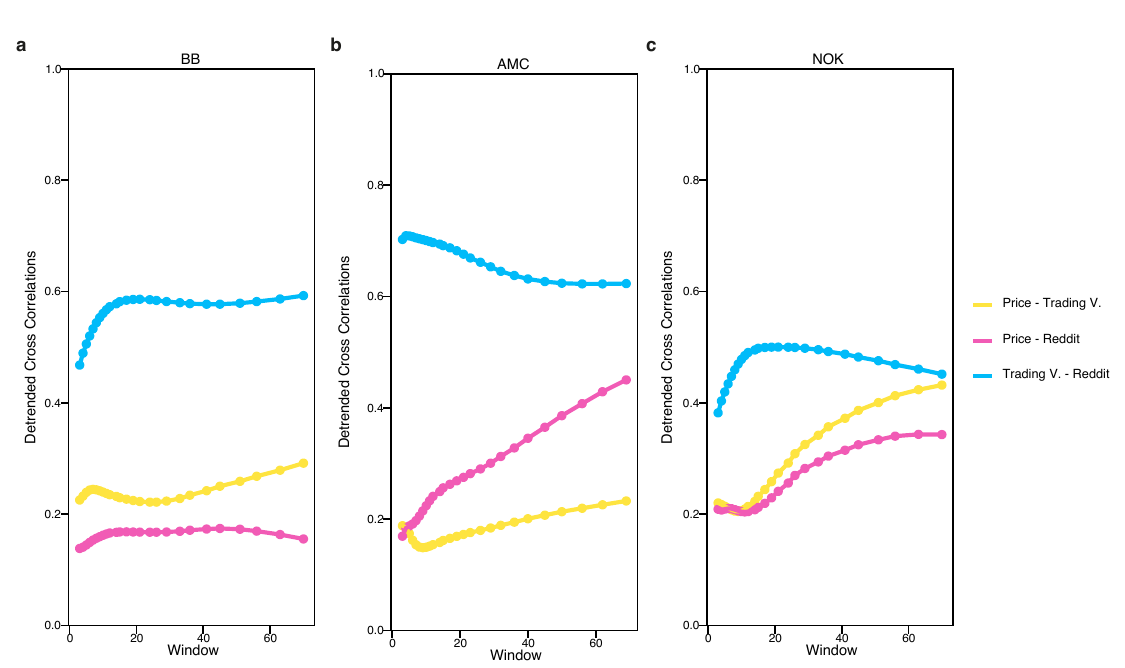}
    \caption{Detrended Cross Correlation coefficients for the following combinations of signals: price / trading volume (light blue), price / Reddit activity (pink), trading volume / Reddit activity (purple), for BB (panel \textbf{a}), AMC (panel \textbf{b}) and NOK (panel \textbf{c}).
    }
    \label{fig_sm_dcca}
\end{figure}

\begin{figure}[h!]
    \centering
    \includegraphics[width=1.0\textwidth]{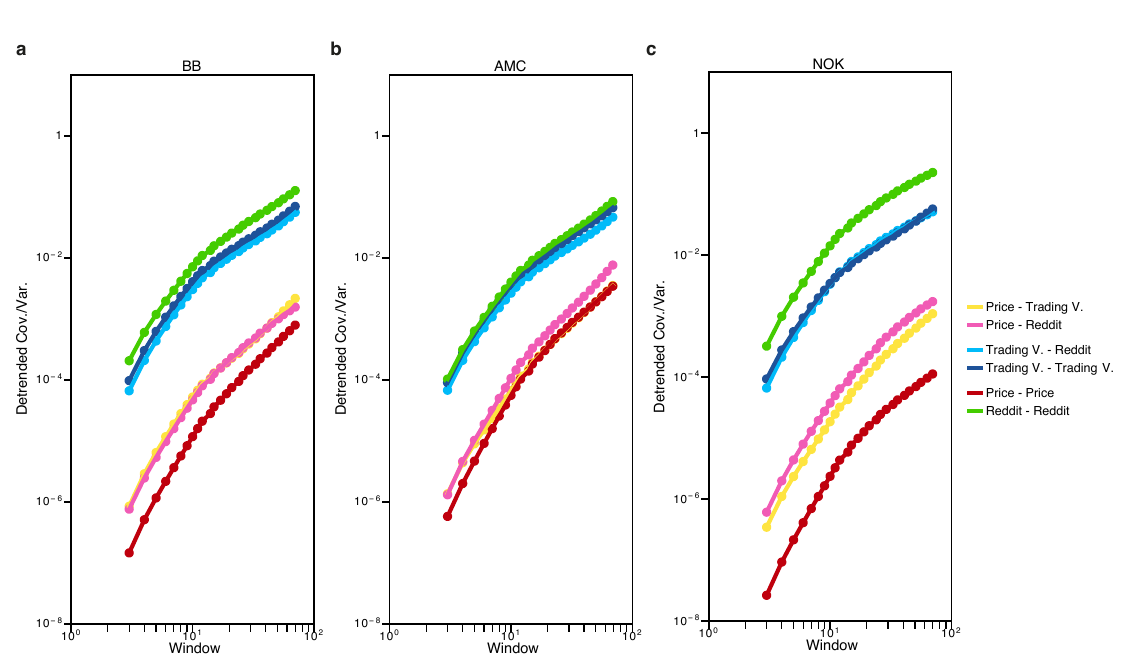}
    \caption{Detrended Variances for each signal: price (red), trading volume (blue) and Reddit activity (green); detrended covariances for their paired combinations (same color-scheme of Figure \ref{fig_sm_dcca}); for BB (panel \textbf{a}), AMC (panel \textbf{b}) and NOK (panel \textbf{c}).
    }
    \label{fig_sm_dcovar}
\end{figure}

As discussed in the main text, we perform the detrended cross-correlation analysis on BB, AMC, NOK. However, in this case, our analysis focuses on data spanning from January 31st to July 1st, 2021.
This time frame is chosen as these stocks were not extensively discussed prior to January, as reported in the main text.
Unlike the GME case reported in the main text, we observe that the cross-correlation coefficients exhibit variation and dependence on the scale $t$ (see Figure \ref{fig_sm_dcca}).
This variability could be attributed to the limited number of time scales considered and it is similar to the variability observed by Podobnik et al. \cite{podobnik_statistical_2011} for the New York and Shanghai Stock Exchange Composite index.
Nevertheless, the power-law relationship of the covariances and variances (see Figure \ref{fig_sm_dcovar}), as well as the fitted scale exponent reported in Table \ref{tab_sm_exponent}, supports the findings reported in the main text.
\begin{table}[h!]
	\centering
    \begin{tabular}{llr}
        \toprule
        Stock &                   Combination &  $\lambda$ \\
        \midrule
            BB &      Price - Trading V. &   0.92 \\
            BB &          Price - Reddit &   0.82 \\
            BB &     Trading V. - Reddit &   0.66 \\
            BB & Trading V. - Trading V. &   0.65 \\
            BB &           Price - Price &   0.99 \\
            BB &         Reddit - Reddit &   0.67 \\
            AMC &      Price - Trading V. &   0.94 \\
            AMC &          Price - Reddit &   0.99 \\
            AMC &     Trading V. - Reddit &   0.66 \\
            AMC & Trading V. - Trading V. &   0.69 \\
            AMC &           Price - Price &   0.93 \\
            AMC &         Reddit - Reddit &   0.71 \\
            NOK &      Price - Trading V. &   0.95 \\
            NOK &          Price - Reddit &   0.89 \\
            NOK &     Trading V. - Reddit &   0.60 \\
            NOK & Trading V. - Trading V. &   0.65 \\
            NOK &           Price - Price &   0.84 \\
            NOK &         Reddit - Reddit &   0.61 \\
            GME &      Price - Trading V. &   0.98 \\
            GME &          Price - Reddit &   0.92 \\
            GME &     Trading V. - Reddit &   0.66 \\
            GME & Trading V. - Trading V. &   0.63 \\
            GME &           Price - Price &   0.93 \\
            GME &         Reddit - Reddit &   0.7 \\
        \bottomrule
    \end{tabular}
    \caption{Scale exponents $\lambda$ of the possible combinations of the detrended covariances and variances for all the stocks. The exponents are fitted starting from window 20.
    }
    \label{tab_sm_exponent}
\end{table}

\begin{figure}[h!]
    \centering
    \includegraphics[width=0.85\textwidth]{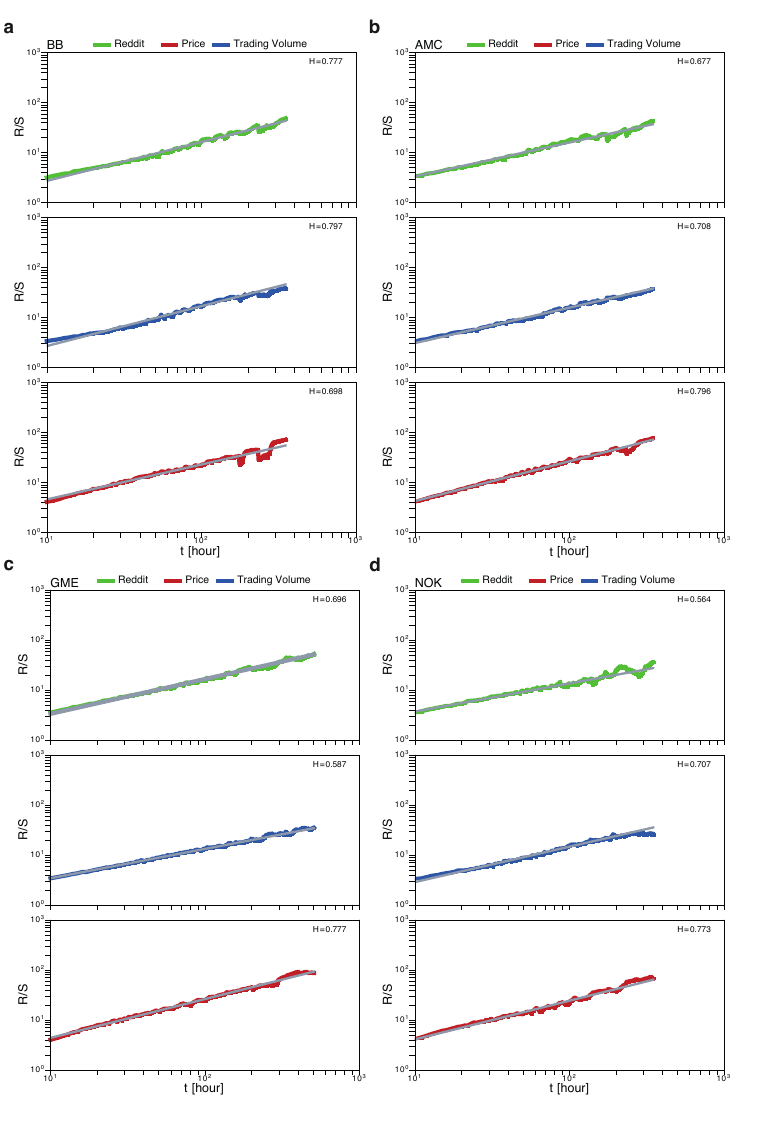}
    \caption{Rescaled range analysis conducted for each signal: price (red), trading volume (blue), and Reddit activity (green). The grey lines represent linear fits, and the corresponding Hurst exponents are provided in the legend.
    In panel \textbf{a}, BB, in panel \textbf{b}, AMC, in panel \textbf{c}, GME, and in panel \textbf{d}, NOK.
    }
    \label{fig_sm_hurst}
\end{figure}

We perform an additional analysis by examining the detrended variance of each signal removing the mean of the signal in each window to further characterise the non-linear dependencies. 
This latter exercise allows us to estimate the Hurst exponent \cite{mantegna_introduction_1999, carbone_time-dependent_2004}.
In this case, instead of considering the variance as a function of the window $t$, we examine the rescaled range $R/S$ \cite{mantegna_introduction_1999}.
The rescaled range is the ratio between the range of the signal, which is the difference between the maximum and minimum values in the cumulative sum of the residuals from the mean, and the standard deviation of the residuals (variance up to a square root).
We consistently observe similar results across the analyzed stocks: the Reddit signals exhibit an almost power-law behavior in the rescaled range as a function of the window $t$, along with a Hurst exponent greater than 0.5 (see Figure \ref{fig_sm_hurst} and Table \ref{tab_sm_hurst}). 
In this case, we represent all windows $t$ since the fluctuations, though less pronounced than the previous analysis, align with common findings in the literature \cite{mantegna_introduction_1999}.
These results suggest a persistent and trending behavior in the Reddit changes, indicating the presence of long-range dependencies in the signal.
To accurately determine the Hurst exponent or the scale exponent, it is recommended to include a broader range of time scales in the analysis -- such as years \cite{mantegna_introduction_1999}.
Our analysis serves as an exercise in exploring possible non-linear dependencies in social media / financial markets dynamics.

\begin{table}[h!]
	\centering
    \begin{tabular}{llr}
    \toprule
    Stock &           Signal &        Hurst Exponent \\
    \midrule
      GME &         Reddit & 0.696 \\
      GME & Trading Volume & 0.587\\
      GME &          Price & 0.777 \\
      NOK &         Reddit & 0.564 \\
      NOK & Trading Volume & 0.707 \\
      NOK &          Price & 0.773 \\
       BB &         Reddit & 0.776 \\
       BB & Trading Volume & 0.797 \\
       BB &          Price & 0.698 \\
      AMC &         Reddit & 0.677 \\
      AMC & Trading Volume & 0.708 \\
      AMC &          Price & 0.796 \\
    \bottomrule
    \end{tabular}
    \caption{Hurst exponent for all stocks and signals. The exponents are fitted from window length 10 to 200.
    }
    \label{tab_sm_hurst}
\end{table}
\pagebreak[2]
\clearpage
\section*{Supplementary Materials, Section 3: Convergent Cross-Mapping}\label{sm_ccm}

To capture and quantify non-linear coupling relations between variables of the same dynamical systems, we employ a statistical technique known as Convergent Cross Mapping (CCM) \cite{sugihara_detecting_2012}.
Similar to Granger Test, the key concept behind CCM is to test whether the knowledge of the past states of variable $x$ can improve predictions of variable $y$ 's future states. 
Unlike Granger Test, CCM does not assume that we can separate the information about $x$ from the rest of the systems, as it is often the case in nonlinear models \cite{runge_detecting_2019,bradley_nonlinear_2015}.
Given a time series $x(t)$, we consider the delay vectors $X_{\tau,D}(t) = (x(t),\dots,x(t-(D-1)\tau))$ and for large enough embedding dimension D (larger than the attractor dimensions) the manifold $M_X$ identified by $X$ reconstructs the original true manifold $M_S$ (or all the topological invariant of the dynamical systems $S$).
Given two time series $x(t)$ and $y(t)$, we assume that these are generated from an unknown high-dimensional dynamical systems $S$. 
Using Takens' theorem \cite{braaksma_numerical_1985,sugihara_nonlinear_1990,rand_detecting_1981}, we can construct from each time series $s$ a manifold $M_s$ and confront at a given time $t$ the points on one reconstructed manifold $(M_X)$ with the other $(M_Y)$.
If $X\to Y$, then information from $X$ is stored in $Y$ and gets embedded in $M_Y$, thus it follows that we can use $M_Y$ to better predict $X$ (the manifold $M_Y$ contain the information of both $X$ and $Y$).

Operationally, given the embedding $D$ and lag $\tau$, we consider the vector \\$\tilde{Y}_{\tau,D}(t)=(y(t),\dots,y(t-(D-1)\tau),x(t))$ and we perform a neighbor search using the first dimensions to predict last dimension, i.e., $x(t)$. 
The correlation $\rho_{YX}$ summarises the results of this prediction \cite{sugihara_detecting_2012}.
CCM requires two criteria to conclude synergistic relation: cross mapping and convergence \cite{sugihara_detecting_2012}.
Cross mapping is linked to the correlation $\rho$ and the stronger the coupling in one (or both) direction, the higher the value of $\rho$, since the prediction error using one manifold is either lower or equal to that of the other.
Convergence refers to improving the cross-mapping with time-series length $T$ (sample size to reconstruct the manifold, referred as library size \cite{sugihara_detecting_2012}) and it is related to Takens' theorem.
The more the data, the more precise is the reconstruction of the attractor, resulting in closer nearest neighbors and higher correlation $\rho$.

Nonetheless, we still need to determine the appropriate embedding dimension $D$ and time lag $\tau$ \cite{braaksma_numerical_1985,sugihara_nonlinear_1990}.
The optimal time lag $\tau$ has been considered as the first minimum of the auto-correlation function of a given time series $x(t)$\cite{bradley_nonlinear_2015}.
With this choice, we add as much new information as possible when we introduce new dimensions to the delay vector.
To compute the optimal embedding dimension $D$, we employ the Cao's method, also known as the approximate false nearest neighbors method \cite{cao_practical_1997}.
False nearest neighbors are points that are close to each other in the lower-dimensional embedding space but far apart in the higher-dimensional space \cite{bradley_nonlinear_2015,camps-valls_discovering_2023}.
The method works as follows. 
First, for a point $i$ in a given dimension $D$, we find its first nearest neighbors $n(i)$.
Next, we calculate the ratio of distances between pairs $(i, n(i))$ when computed in dimension $D$ and dimension $D+1$.
Then, we calculate the average ratio for all such pairs, defining it as $a_D$.
The number of false nearest neighbors is monotonically decreasing with $D$ and therefore the ratio
$E_1 = a_{D+1}/a_{D}$ converges slowly to 1.
The method assumes one already has an optimal time lag $\tau$, and then compute the statistics $E_1$ as function of the dimension $D$.
The optimal embedding dimension is the one corresponding to the knee point of the statistics $E_1$, as it balances between capturing the underlying dynamics and avoiding excessive sparse points.

\begin{figure}[h!]
    \centering
    \includegraphics[width=1.0\textwidth]{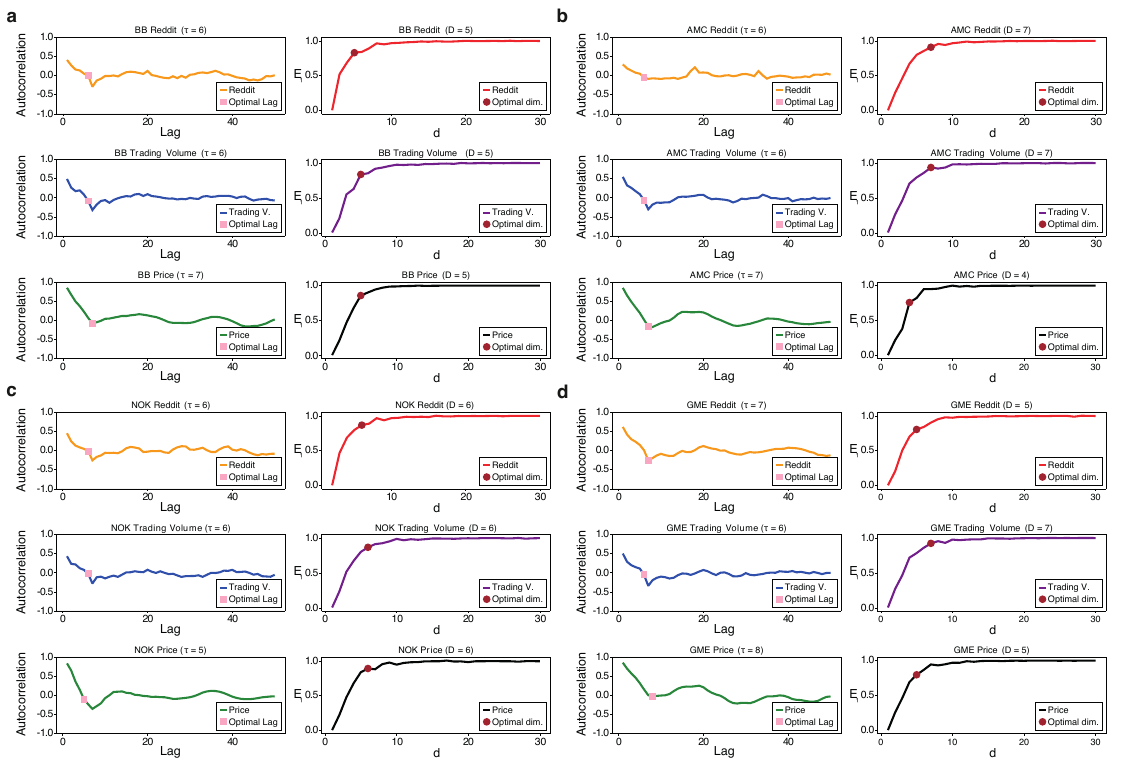}
    \caption{\textbf{Optimal lag and dimension for all the stocks.}
    Autocorrelation (Left) and $E_1$ statistics (Right) for all the signals (each sub-panel) and all the stocks (each panel). The markers identified the optimal lag (square) and the optimal dimension (circle).
    In panel \textbf{a}, BB, in panel \textbf{b}, AMC, in panel \textbf{c}, NOK, and in panel \textbf{d}, GME.
    Autocorrelation Reddit (orange), trading volume (blue), price (green).
    $E_1$ statistics Reddit (red), trading volume (purple), price (black).
    }
    \label{fig_sm_embed_delay}
\end{figure}

\begin{figure}[h!]
    \centering
    \includegraphics[width=1.0\textwidth]{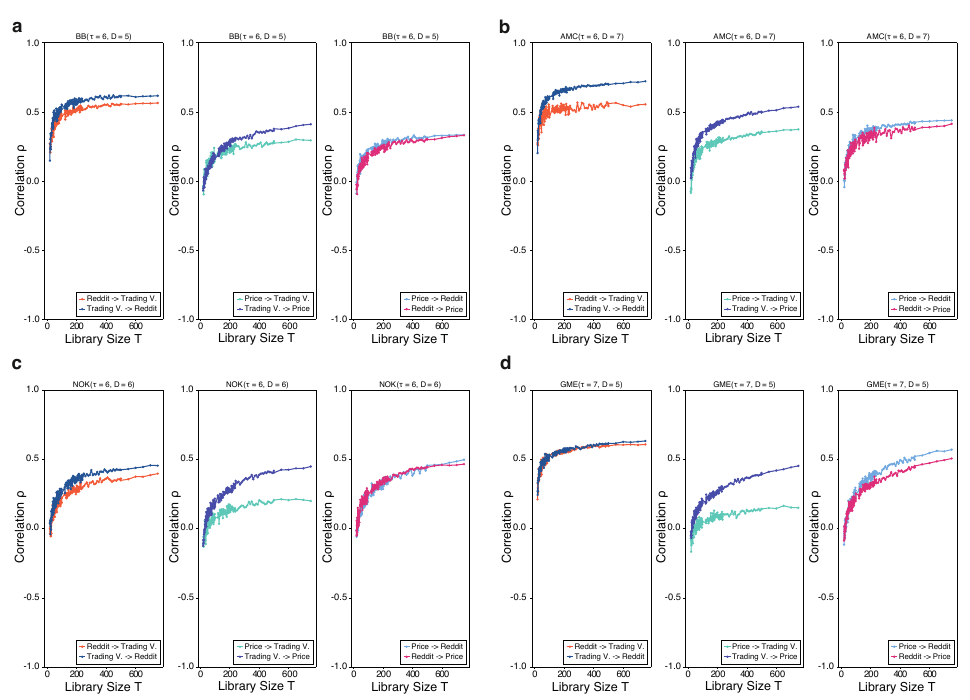}
    \caption{
    \textbf{Convergent Cross-Mapping for all the stocks.}
    Correlation $(\rho)$ as a function of the length of time series $T$ for optimal lag $\tau$ and optimal dimension $D$ for different combinations of signals: Reddit-to-trading volume (orange, left panel), trading volume-to-Reddit (blue, left panel),
    trading volume-to-price (purple, center), price-to-trading volume (light green, center), price-to-Reddit (light blue, right panel), Reddit-to-price (pink, right panel).
    In panel \textbf{a}, BB $(\tau = 6, D = 5)$, in panel \textbf{b}, AMC $(\tau = 6, D = 7)$, in panel \textbf{c}, NOK $(\tau = 6, D = 6)$, and in panel \textbf{d}, GME $(\tau = 7, D = 5)$.
    }
    \label{fig_sm_cm}
\end{figure}

\begin{figure}[h!]
    \centering
    \includegraphics[width=0.85\textwidth]{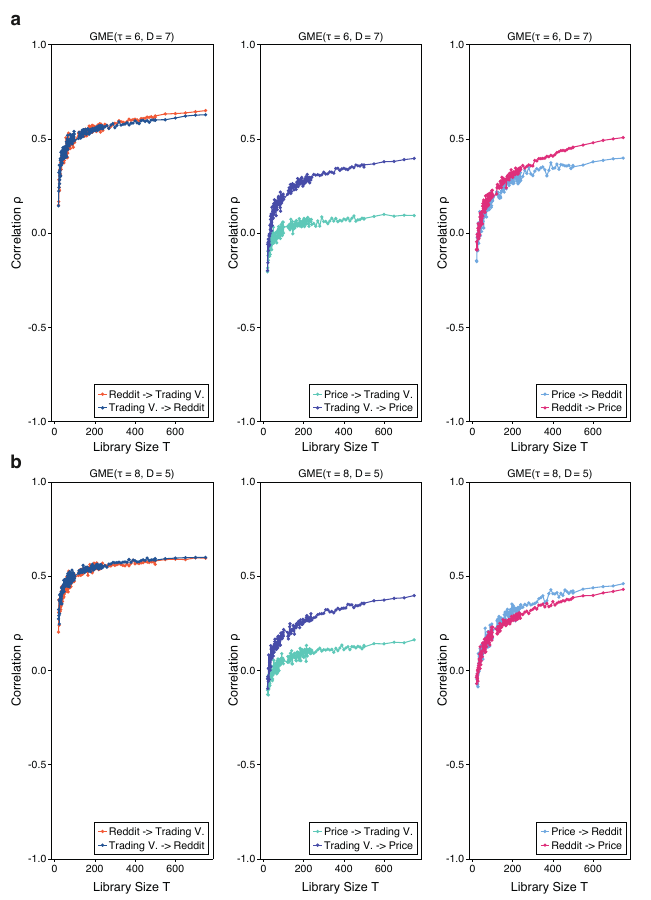}
    \caption{\textbf{Convergent Cross-Mapping for GME for different optimal lag and dimension.}
    Correlation $\rho$ as a function of length of time series $T$ for different combinations of GME signals: Reddit-to-trading volume (orange, left panel), trading volume-to-Reddit (blue, left panel), trading volume-to-price (purple, center), price-to-trading volume (light green, center), price-to-Reddit (light blue, right panel), Reddit-to-price (pink, right panel).
    In panel \textbf{a} optimal lag $\tau = 6$ and optimal dimension $D = 7$, and in panel \textbf{b} optimal lag $\tau = 8$ and optimal dimension $D = 5$. We observe consistent results with Figure \ref{fig_sm_cm}.
    }
    \label{fig_sm_cm_sup}
\end{figure}

Following the Granger analysis and the approach taken by Stavroglou et al. \cite{stavroglou_hidden_2019}, we have considered the hourly return of the Reddit, trading volume and price signals. 
Figure \ref{fig_sm_embed_delay} shows the autocorrelation and the $E_1$ statistics for all stocks signals. 
Notably, we observe that the optimal lag falls within the range of $6$ to $8$, and the optimal dimension lies within the range of $5$ to $7$. 
Furthermore, we have found that for every stock, there are at least two combinations of the three signals that have the same optimal lag and dimension.
We have specifically selected this combination for computing the correlation $\rho$ in Convergent Cross Mapping.
Following Luo et al. \cite{luo_questionable_2015}, we have computed the average correlation $\rho$ on contiguous time series windows of length $T$ with a randomly selected starting point (100 repetitions).
This approach reduces the influence of noise for small windows, thereby resulting in a smoother convergence\cite{sugihara_detecting_2012,luo_questionable_2015}.
Figure \ref{fig_sm_cm} shows the correlation $\rho$ for all the combinations of the signals.
The changes in trading volume and Reddit conversations are mutually coupled, as the correlation coefficients in both directions are higher than the others.
Moreover, there is a marginally greater influence of the trading volume on Reddit conversations (negligible for GME).
This result should not surprise since platforms like Reddit tend to influence the stock market primarily during specific events and with short-term impacts \cite{vassallo_tale_2022}.
The stock market serves as the primary source of information, aggregating information from multiple channels, thus it is more likely that trading volume changes predate Reddit changes.
Furthermore, the findings of the price indicate that trading volume assumes a dominant role -- large variations of trading volume were essential to trigger the short squeezes.
Figure \ref{fig_sm_cm_sup} displays the correlation cross-mapping for the GME case using various combinations of optimal dimensions and lags, and we note no variations in the results.
These results corroborate that changes in trading volume are more tied to the Reddit discussions, rather than to price movements. 
\pagebreak[2]
\clearpage

\section*{Supplementary Materials, Section 4: Screenshot analysis}\label{sm_screenshot}

As detailed in the main text, we employ computer vision techniques to extract actual financial investments within the WSB community. 
Indeed users in the WSB community typically share screenshots of their open positions. 
Lucchini et al.  \cite{lucchini_reddit_2022} quantified the WSB community's commitment to GME by parsing these screenshots using a combination of text analysis and computer vision techniques.
We rely on this previous analysis and consider the dataset of 78468 screenshots as our initial dataset.
We note that these screenshots exhibit arbitrary structures, as WSB users do not use a specific app and screenshots are handmade, resulting in a lack of standardized format.
Through manual inspection of a sample of the screenshots, we identify a set of keywords that proved useful in narrowing down the selection to those representing actual investments (6525 screenshots), as discussed in the main text.
Our extraction technique, for instance, does not accommodate the possibility of multiple orders in a single screenshot, as we extract only the highest value.

\begin{figure}[b!]
    \centering
    \includegraphics[width=1.0\textwidth]{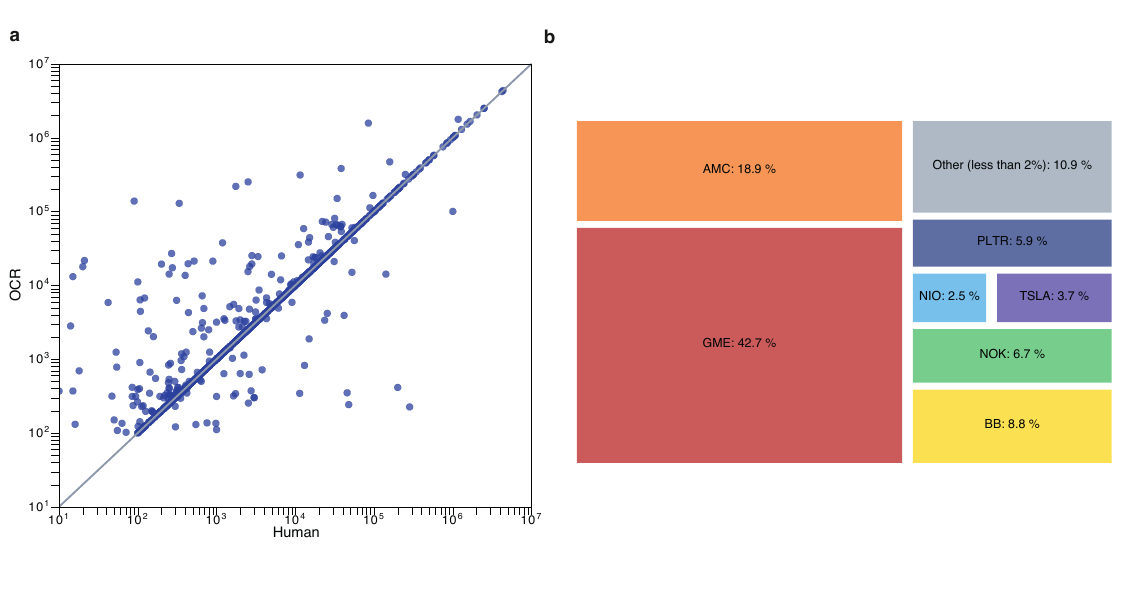}
    \caption{\textbf{Comparison between OCR and human extraction and stocks rank.}
    \textbf{a}) Scatter plot displaying the comparison between position values extracted using OCR and human visual inspection.
    \textbf{b}) Treemap rank chart showing the occurrences of stock tickers in screenshots.
    }
    \label{fig_sm_screen}
\end{figure}

Given this arbitrariness, we consider a sample of 1000 screenshot and validate the extraction through manual inspection.
Figure \ref{fig_sm_screen}\textbf{a} displays the comparison between OCR technique and manual inspection, with an accuracy of 0.85 and an RMSE of 0.47, where RMSE is computed using log-transformed data and accuracy is determined through a binary classification method with a threshold of 0.05 for correctness.
We perform an additional exercise using all the dataset \cite{lucchini_reddit_2022} and extract all stocks ticker symbols in a given screenshot.
Figure \ref{fig_sm_screen}\textbf{b} summarises the popularity of a given stock in the screenshots, notably observing the same stock ranking as reported in the main text using ticker occurrences in posts/comments.
Table \ref{tab_sm_co_screens_stocks} details the number of co-occurrences of two stocks in a screenshot.
Notably, we observe that the stocks AMC, BB, NOK and GME exhibit a clustering pattern, while PLTR, TSLA and NIO show comparatively fewer connections, primarily appearing in conjunction with GME.

\begin{table}[h!]
	\centering
    \begin{tabular}{llr}
    \toprule
    Stock A & Stock B &  Number of Screenshots \\
    \midrule
       AMC &   PLTR &      56 \\
       GME &    NIO &      62 \\
       NOK &   PLTR &      62 \\
       NIO &   PLTR &      65 \\
       AMC &    NIO &      70 \\
       NIO &   TSLA &      80 \\
      PLTR &   TSLA &      81 \\
       AMC &   TSLA &      98 \\
        BB &   PLTR &     132 \\
       GME &   TSLA &     148 \\
       GME &   PLTR &     183 \\
        BB &    NOK &     457 \\
       GME &    NOK &     484 \\
       AMC &     BB &     648 \\
       AMC &    NOK &     698 \\
        BB &    GME &     719 \\
       AMC &    GME &    1773 \\
    \bottomrule
    \end{tabular}
    \caption{Cumulative number of screenshots for stock ticker pairs.
    } 
\label{tab_sm_co_screens_stocks}
\end{table}
\pagebreak[2]
\clearpage
\section*{Supplementary Materials, Section 5: Robustness checks on Granger Test}\label{sm_gc}
\subsection*{Granger Test: stationarity tests}
\begin{table}[h!]
    \centering
    \begin{adjustbox}{center}
    \resizebox{0.4\textwidth}{!}{
        \begin{tabular}{l|cc|cc|cc|}
        &  \multicolumn{2}{|c|}{Twitter} & \multicolumn{2}{|c|}{Trading Volume} &  \multicolumn{2}{|c|}{Reddit} \\
        Date &  t &  p-value &  t &  p-value &  t &  p-value\\
        \midrule
        2020-12-17 &       -4.296 &              0.000 &              -4.299 &                     0.000 &      -4.429 &             0.000 \\
        2020-12-18 &       -4.251 &              0.001 &              -4.236 &                     0.001 &      -4.504 &             0.000 \\
        2020-12-19 &       -4.894 &              0.000 &              -4.233 &                     0.001 &      -4.405 &             0.000 \\
        2020-12-20 &       -4.894 &              0.000 &              -4.233 &                     0.001 &      -4.405 &             0.000 \\
        2020-12-21 &       -4.894 &              0.000 &              -4.233 &                     0.001 &      -4.405 &             0.000 \\
        2020-12-22 &       -5.000 &              0.000 &              -4.168 &                     0.001 &      -4.350 &             0.000 \\
        2020-12-23 &       -4.627 &              0.000 &              -5.939 &                     0.000 &      -5.031 &             0.000 \\
        2020-12-24 &       -5.182 &              0.000 &              -4.987 &                     0.000 &      -5.315 &             0.000 \\
        2020-12-25 &       -5.149 &              0.000 &              -4.134 &                     0.001 &      -4.947 &             0.000 \\
        2020-12-26 &       -4.853 &              0.000 &              -3.853 &                     0.002 &      -4.601 &             0.000 \\
        2020-12-27 &       -4.853 &              0.000 &              -3.853 &                     0.002 &      -4.601 &             0.000 \\
        2020-12-28 &       -4.853 &              0.000 &              -3.853 &                     0.002 &      -4.601 &             0.000 \\
        2020-12-29 &       -4.751 &              0.000 &              -3.880 &                     0.002 &      -4.666 &             0.000 \\
        2020-12-30 &       -4.729 &              0.000 &              -4.091 &                     0.001 &      -4.712 &             0.000 \\
        2020-12-31 &       -1.841 &              0.360 &              -4.280 &                     0.000 &      -5.348 &             0.000 \\
        2021-01-01 &       -3.202 &              0.020 &              -4.657 &                     0.000 &      -5.688 &             0.000 \\
        2021-01-02 &       -2.931 &              0.042 &              -4.244 &                     0.001 &      -5.259 &             0.000 \\
        2021-01-03 &       -2.931 &              0.042 &              -4.244 &                     0.001 &      -5.259 &             0.000 \\
        2021-01-04 &       -2.931 &              0.042 &              -4.244 &                     0.001 &      -5.259 &             0.000 \\
        2021-01-05 &       -2.938 &              0.041 &              -4.258 &                     0.001 &      -5.288 &             0.000 \\
        2021-01-06 &       -2.698 &              0.074 &              -4.552 &                     0.000 &      -4.940 &             0.000 \\
        2021-01-07 &       -2.715 &              0.072 &              -4.689 &                     0.000 &      -4.820 &             0.000 \\
        2021-01-08 &       -2.788 &              0.060 &              -5.738 &                     0.000 &      -5.764 &             0.000 \\
        2021-01-09 &       -2.931 &              0.042 &              -6.234 &                     0.000 &      -6.076 &             0.000 \\
        2021-01-10 &       -2.931 &              0.042 &              -6.234 &                     0.000 &      -6.076 &             0.000 \\
        2021-01-11 &       -2.931 &              0.042 &              -6.234 &                     0.000 &      -6.076 &             0.000 \\
        2021-01-12 &       -3.698 &              0.004 &              -6.108 &                     0.000 &      -5.725 &             0.000 \\
        2021-01-13 &       -3.815 &              0.003 &              -5.685 &                     0.000 &      -4.988 &             0.000 \\
        2021-01-14 &       -1.185 &              0.680 &              -4.290 &                     0.000 &      -4.738 &             0.000 \\
        2021-01-15 &       -5.310 &              0.000 &              -4.294 &                     0.000 &      -4.785 &             0.000 \\
        2021-01-16 &       -5.680 &              0.000 &              -4.425 &                     0.000 &      -5.053 &             0.000 \\
        2021-01-17 &       -5.680 &              0.000 &              -4.425 &                     0.000 &      -5.053 &             0.000 \\
        2021-01-18 &       -5.680 &              0.000 &              -4.425 &                     0.000 &      -5.053 &             0.000 \\
        2021-01-19 &       -5.346 &              0.000 &              -4.167 &                     0.001 &      -4.937 &             0.000 \\
        2021-01-20 &       -5.225 &              0.000 &              -4.147 &                     0.001 &      -4.752 &             0.000 \\
        2021-01-21 &       -5.104 &              0.000 &              -3.747 &                     0.004 &      -4.792 &             0.000 \\
        2021-01-22 &       -5.362 &              0.000 &              -4.006 &                     0.001 &      -4.843 &             0.000 \\
        2021-01-23 &       -5.393 &              0.000 &              -3.965 &                     0.002 &      -4.949 &             0.000 \\
        2021-01-24 &       -5.393 &              0.000 &              -3.965 &                     0.002 &      -4.949 &             0.000 \\
        2021-01-25 &       -5.393 &              0.000 &              -3.965 &                     0.002 &      -4.949 &             0.000 \\
        2021-01-26 &       -4.094 &              0.001 &              -4.004 &                     0.001 &      -5.068 &             0.000 \\
        2021-01-27 &       -4.972 &              0.000 &              -3.601 &                     0.006 &     -10.878 &             0.000 \\
        2021-01-28 &       -4.298 &              0.000 &              -4.008 &                     0.001 &      -3.058 &             0.030 \\
        2021-01-29 &       -4.189 &              0.001 &              -4.101 &                     0.001 &      -2.560 &             0.102 \\
        2021-01-30 &       -3.637 &              0.005 &              -4.127 &                     0.001 &      -2.636 &             0.086 \\
        2021-01-31 &       -3.637 &              0.005 &              -4.127 &                     0.001 &      -2.636 &             0.086 \\
        2021-02-01 &       -3.637 &              0.005 &              -4.127 &                     0.001 &      -2.636 &             0.086 \\
        2021-02-02 &       -3.254 &              0.017 &              -4.352 &                     0.000 &      -2.881 &             0.048 \\
        2021-02-03 &       -3.864 &              0.002 &              -4.358 &                     0.000 &      -2.767 &             0.063 \\
        2021-02-04 &       -3.460 &              0.009 &              -3.839 &                     0.003 &      -2.696 &             0.075 \\
        2021-02-05 &       -3.702 &              0.004 &              -3.995 &                     0.001 &      -2.989 &             0.036 \\
        2021-02-06 &       -3.690 &              0.004 &              -4.153 &                     0.001 &      -3.051 &             0.030 \\
        2021-02-07 &       -3.690 &              0.004 &              -4.153 &                     0.001 &      -3.051 &             0.030 \\
        2021-02-08 &       -3.690 &              0.004 &              -4.153 &                     0.001 &      -3.051 &             0.030 \\
        2021-02-09 &       -3.001 &              0.035 &              -3.754 &                     0.003 &      -2.262 &             0.184 \\
        2021-02-10 &       -2.733 &              0.069 &              -3.412 &                     0.011 &      -2.268 &             0.182 \\
        2021-02-11 &       -2.702 &              0.074 &              -3.446 &                     0.009 &      -2.765 &             0.063 \\
        2021-02-12 &       -3.432 &              0.010 &              -3.196 &                     0.020 &      -2.123 &             0.235 \\
        2021-02-13 &       -3.045 &              0.031 &              -3.006 &                     0.034 &      -1.854 &             0.354 \\
        \bottomrule
        \end{tabular}
    }
    \end{adjustbox}
    \caption{Results of the Augmented Dickey-Fuller test conducted on the Twitter, Trading Volume, and Reddit signals to assess stationarity. 
    The test statistics (t) and corresponding p-values (rounded to the third decimal) are reported for each signal in various time windows, identified by the Date column. 
    Each row represents a separate test conducted on the respective signal within the specified time window.
    }
    \label{tab_sm_ stationarity}
\end{table}
\pagebreak[2]
\clearpage

\subsection*{Granger Test: p-values and coefficient of the Autoregressive Model}
\begin{table}[h!]
    \centering
    \begin{adjustbox}{center}
    \resizebox{0.99\textwidth}{!}{
    \begin{tabular}{lcccccc}
    \toprule
    Date &  {Reddit $->$ Trading Volume} & {Trading Volume $->$ Reddit} & {Reddit $->$ Twitter} & {Twitter $->$ Reddit} & {Twitter $->$ Trading Volume} &  {Trading Volume $->$ Twitter}\\
    \midrule
    2021-01-02 & 0.909 & 0.041 & 0.687 & 0.742 & 0.934 & 0.197 \\
    2021-01-03 & 0.909 & 0.041 & 0.687 & 0.742 & 0.934 & 0.197 \\
    2021-01-04 & 0.909 & 0.041 & 0.687 & 0.742 & 0.934 & 0.197 \\
    2021-01-05 & 0.606 & 0.07 & 0.651 & 0.804 & 0.998 & 0.176 \\
    2021-01-06 & 0.488 & 0.757 & 0.846 & 0.751 & 0.603 & 0.225 \\
    2021-01-07 & 0.458 & 0.815 & 0.963 & 0.81 & 0.62 & 0.296 \\
    2021-01-08 & 0.606 & 0.698 & 0.716 & 0.734 & 0.549 & 0.282 \\
    2021-01-09 & 0.479 & 0.602 & 0.742 & 0.876 & 0.578 & 0.266 \\
    2021-01-10 & 0.479 & 0.602 & 0.742 & 0.876 & 0.578 & 0.266 \\
    2021-01-11 & 0.479 & 0.602 & 0.742 & 0.876 & 0.578 & 0.266 \\
    2021-01-12 & 0.462 & 0.472 & 0.833 & 0.756 & 0.983 & 0.3 \\
    2021-01-13 & 0.507 & 0.855 & 0.67 & 0.589 & 0.987 & 0.226 \\
    2021-01-14 & 0 & 0.001 & 0.703 & 0.327 & 0.43 & 0.636 \\
    2021-01-15 & 0 & 0.001 & 0.634 & 0.531 & 0.618 & 0.769 \\
    2021-01-16 & 0 & 0.003 & 0.54 & 0.575 & 0.811 & 0.521 \\
    2021-01-17 & 0 & 0.003 & 0.54 & 0.575 & 0.811 & 0.521 \\
    2021-01-18 & 0 & 0.003 & 0.54 & 0.575 & 0.811 & 0.521 \\
    2021-01-19 & 0 & 0.006 & 0.524 & 0.611 & 0.884 & 0.497 \\
    2021-01-20 & 0 & 0.023 & 0.571 & 0.675 & 0.958 & 0.603 \\
    2021-01-21 & 0 & 0.033 & 0.487 & 0.714 & 0.877 & 0.499 \\
    2021-01-22 & 0 & 0.03 & 0.42 & 0.681 & 0.93 & 0.461 \\
    2021-01-23 & 0 & 0.045 & 0.515 & 0.64 & 0.816 & 0.649 \\
    2021-01-24 & 0 & 0.045 & 0.515 & 0.64 & 0.816 & 0.649 \\
    2021-01-25 & 0 & 0.045 & 0.515 & 0.64 & 0.816 & 0.649 \\
    2021-01-26 & 0.002 & 0.086 & 0.579 & 0.722 & 0.742 & 0.797 \\
    2021-01-27 & 0.001 & 0.036 & 0.769 & 0.92 & 0.938 & 0.91 \\
    2021-01-28 & 0.339 & 0.663 & 0.067 & 0.518 & 0.741 & 0.865 \\
    2021-01-29 & 0.26 & 0.296 & 0.1 & 0.487 & 0.611 & 0.758 \\
    2021-01-30 & 0.346 & 0.321 & 0.084 & 0.764 & 0.457 & 0.946 \\
    2021-01-31 & 0.346 & 0.321 & 0.084 & 0.764 & 0.457 & 0.946 \\
    2021-02-01 & 0.346 & 0.321 & 0.084 & 0.764 & 0.457 & 0.946 \\
    2021-02-02 & 0.405 & 0.31 & 0.126 & 0.787 & 0.848 & 0.571 \\
    2021-02-03 & 0.189 & 0.474 & 0.117 & 0.919 & 0.369 & 0.797 \\
    2021-02-04 & 0.342 & 0.504 & 0.077 & 0.967 & 0.416 & 0.603 \\
    2021-02-05 & 0.308 & 0.501 & 0.127 & 0.888 & 0.449 & 0.714 \\
    2021-02-06 & 0.183 & 0.602 & 0.095 & 0.92 & 0.482 & 0.819 \\
    2021-02-07 & 0.183 & 0.602 & 0.095 & 0.92 & 0.482 & 0.819 \\
    2021-02-08 & 0.183 & 0.602 & 0.095 & 0.92 & 0.482 & 0.819 \\
    2021-02-09 & 0.225 & 0.739 & 0.118 & 0.986 & 0.287 & 0.796 \\
    2021-02-10 & 0.29 & 0.795 & 0.425 & 0.842 & 0.215 & 0.907 \\
    2021-02-11 & 0.657 & 0.749 & 0.999 & 0.194 & 0.501 & 0.938 \\
    2021-02-12 & 0.239 & 0.783 & 0.821 & 0.191 & 0.805 & 0.521 \\
    2021-02-13 & 0.408 & 0.817 & 0.932 & 0.302 & 0.641 & 0.451 \\
    \bottomrule
    \end{tabular}
    }
    \end{adjustbox}
    \caption{Results of the Granger test between Reddit, Trading Volume, and Twitter signals (GME).
    The table reports the p-values (rounded to the third decimal) for the different directions across time windows identified by the Date column.
    Each row represents a separate test conducted for a specific direction: Reddit anticipating Trading Volume, Trading Volume anticipating Reddit, Reddit anticipating Twitter, Twitter anticipating Reddit, Twitter anticipating Trading Volume, and Trading Volume anticipating Twitter.
    }
    \label{tab_sm_granger_pvalue}
\end{table}

\begin{figure}[!h]
    \centering
    \includegraphics[width=0.5\textwidth]{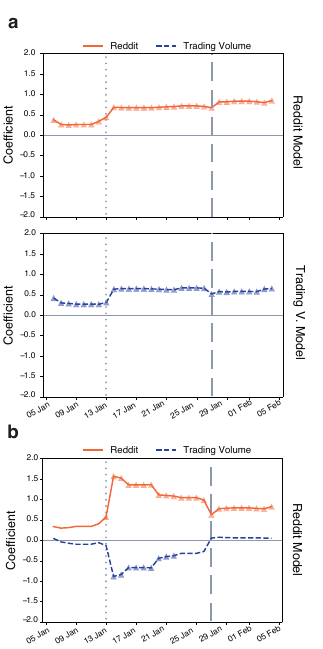}
    \caption{Coefficients of the autoregressive model capturing the relationship between signals over time.
    In panel \textbf{a}, the coefficient is computed for the restricted model (with only one signal): Reddit (top) and Trading Volume (bottom). In panel \textbf{b}, the coefficients are computed for the full model predicting Reddit.
    Each point represents the coefficients over time series spanning from 5 January 2021 to 5 February 2021.
    The vertical gray dotted line indicates the initiation of the Action phase (13 January 2021), whereas the vertical gray dashed line corresponds to the tweet by Elon Musk (27 January 2021). 
    The triangles refer to statistical significant test (p-value $p<0.1$) within the window.
    }
    \label{fig_sm_granger_coeff}
\end{figure}
\pagebreak[2]
\clearpage
\subsection*{Granger Test and Multivariate Vector Autoregression model with a lag of 7}

\begin{figure}[!b]
    \centering
    \includegraphics[width=0.99\textwidth]{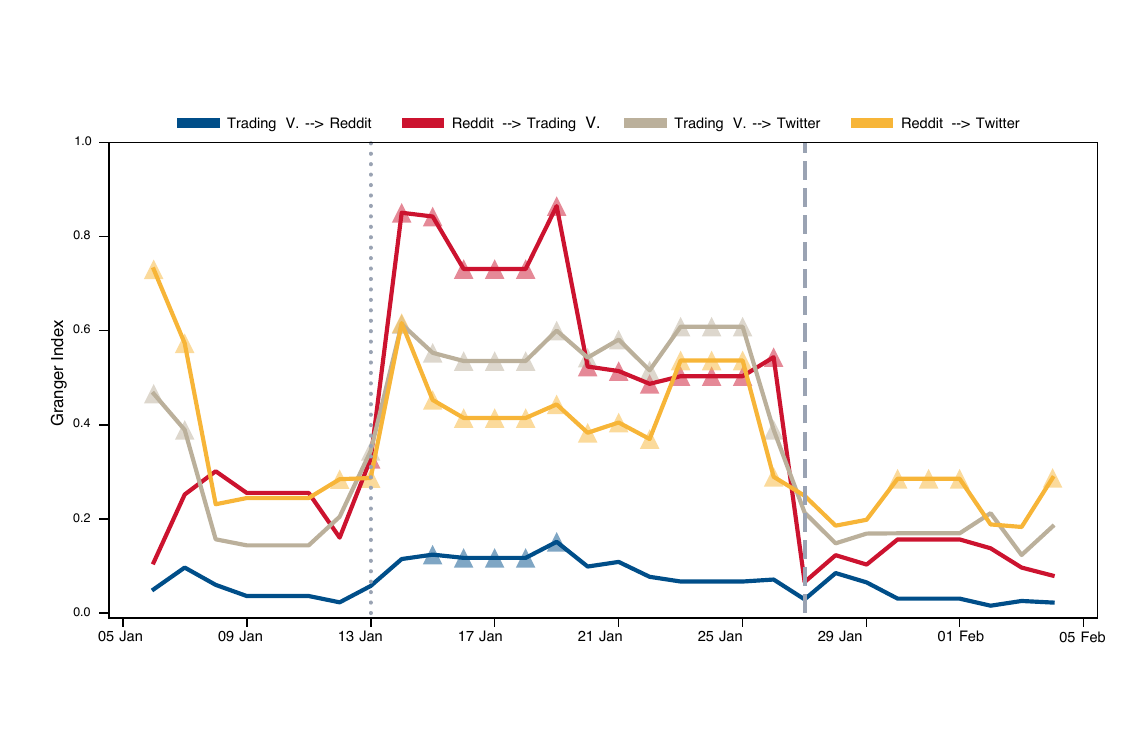}
    \caption{Granger Index capturing the predictive power of a signal on another (with a lag of 1 day) for the following pairs: Trading Volume-to-Reddit (blue), Reddit-to-Trading Volume (red), Trading Volume-to-Twitter (gray) and Reddit-to-Twitter (yellow).
    The vertical gray dotted line indicates the initiation of the \emph{Action} phase (13 January 2021), whereas the vertical gray dashed line corresponds to the Tweet by Elon Musk (27 January 2021) that marks the start of the \emph{Visibility} phase.
    The triangles correspond to statistical significant values (p-values $<0.1$).
    }
    \label{fig_sm_gi_lag_7}
\end{figure}

Figure \ref{fig_sm_gi_lag_7} displays the Granger Index for various combinations of GME signals at lag 7 (approximately 1 day), both when it is significant (p-value $p<0.1$) and when it is not. 
The Granger Index in the Reddit-to-Trading Volume direction becomes significant during the Action phase, similar to the analysis at lag 1 reported in the main text. 
Notably, the magnitude of the Granger Index at lag 7 is higher than at lag 1, and the Reddit-to-Trading Volume direction shows the strongest relationship among all possible combinations. 
Furthermore, we observe a significant Granger Index in the Reddit-to-Twitter direction from the Action and Visibility phases, and in the Trading Volume-to-Twitter direction for the Action phase only. 

\begin{figure}[h!]
    \centering
    \includegraphics[width=0.85\textwidth]{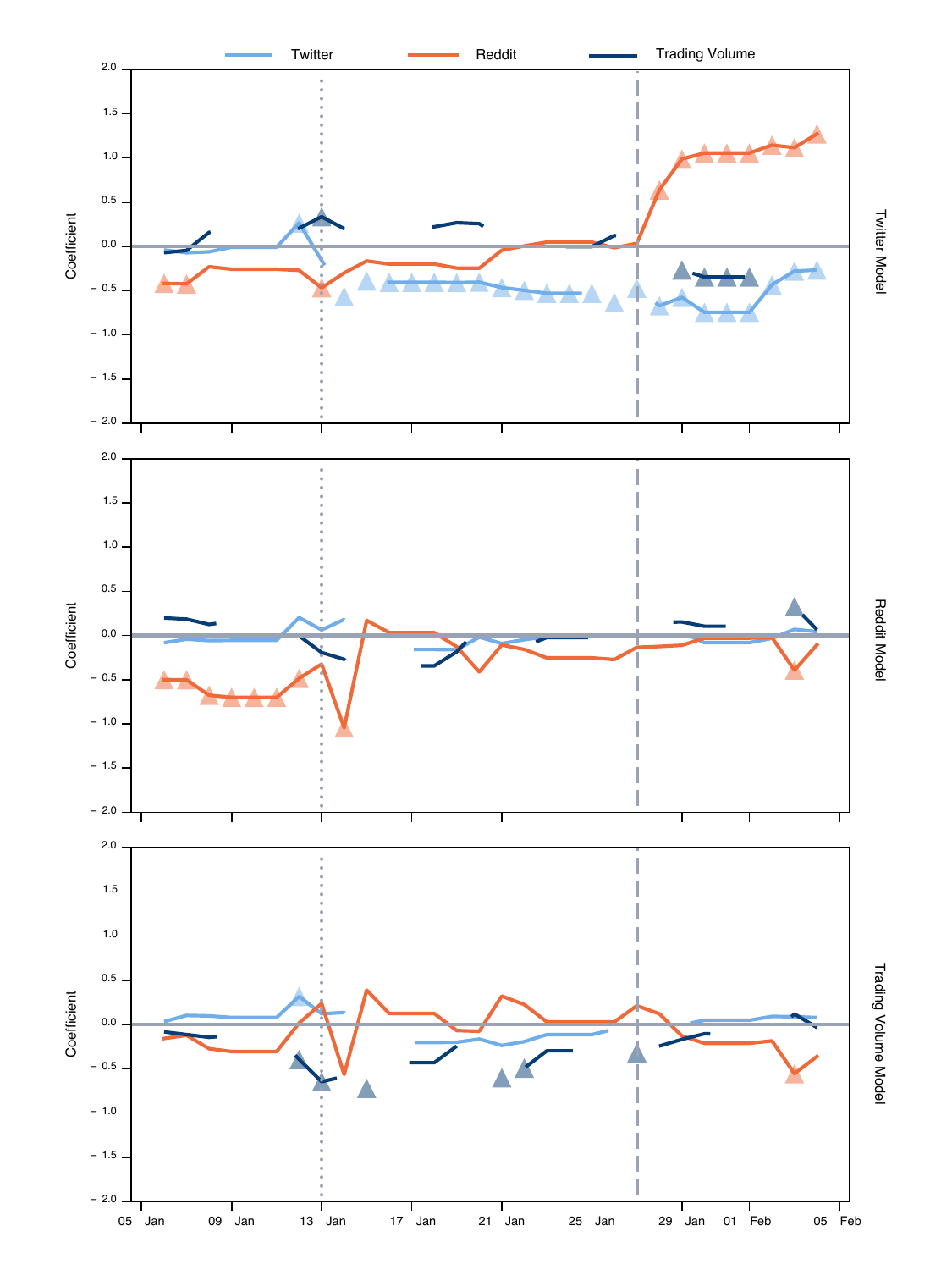}
    \caption{Each panel shows the coefficients of the multivariate autoregressive model at lag 7 predicting Twitter (top), Reddit (middle), and Trading Volume (bottom) activities, from antecedent values of Reddit (orange), Twitter (lightblue), and Trading Volume (blue). 
    The vertical gray dotted line indicates the initiation of the Action phase (13 January 2021), whereas the vertical gray dashed line corresponds to the tweet by Elon Musk (27 January 2021). 
    The triangles refer to statistical significant test (p-value $p<0.1$) within the 15-days window.
    }
    \label{fig_sm_coeff_lag_7}
\end{figure}

However, when we consider the Multivariate Vector Autoregression model, we find different results compared to those reported in the main text.
Figure \ref{fig_sm_coeff_lag_7} displays the different coefficients for each signal in various models at lag 7, both when it is significant (p-value $p<0.1$) and when it is not. 
We observe that the coefficients are consistently close to zero and not statistically significant, except for the Reddit-to-Twitter coefficient in the upper panel, which becomes significant and positive during the Visibility phase.
The coefficients at lag 1 are the same of the analysis reported in the main text. 
Overall, these results corroborate that the shortest time scales are of more significance in the social-media/market dynamics. 
Furthermore, with the inclusion of lags, the model's error diminishes as the information content increases \cite{box_time_2008}.

\pagebreak[2]
\clearpage
\pagebreak[2]
\clearpage
\subsection*{Granger Test and Multivariate Vector Autoregression model for different window length}

\begin{figure}[!b]
    \centering
    \includegraphics[width=1.0\textwidth]{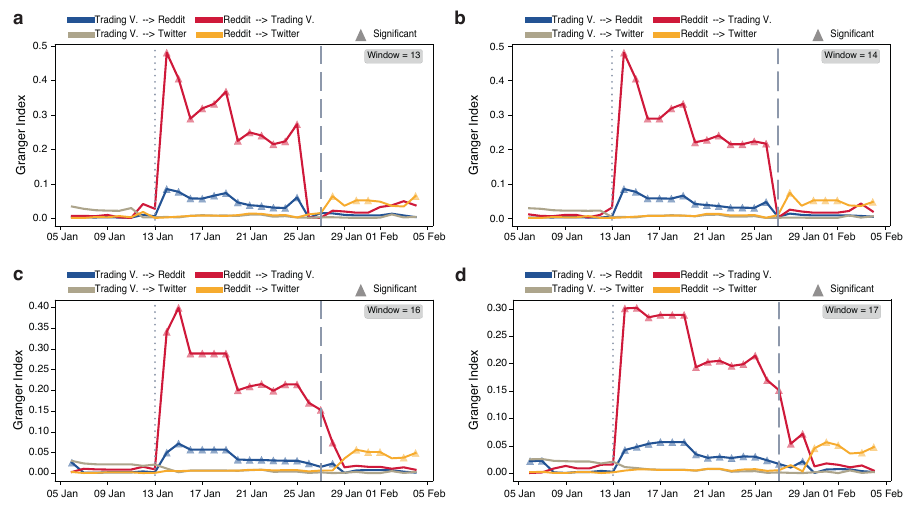}
    \caption{Granger Index capturing the predictive power of a signal on another (with a lag of 1 hour) for the following pairs: Trading Volume-to-Reddit (blue), Reddit-to-Trading Volume (red), Trading Volume-to-Twitter (gray) and Reddit-to-Twitter (yellow).
    Each point is computed considering time series spanning the $w$ preceding days.
    In panel \textbf{a} $w = 13$, in panel \textbf{b} $w = 14$, in panel \textbf{c} $w = 16$ and in panel \textbf{d} $w = 17$.
    The vertical gray dotted line indicates the initiation of the Action phase (13 January 2021), whereas the vertical gray dashed line corresponds to the tweet by Elon Musk (27 January 2021). 
    The triangles refer to statistical significant test (p-value $p<0.1$) within the window.
    }
    \label{fig_sm_gi_window}
\end{figure}

We replicate the analysis outlined in the main text, specifically employing Granger Test and the multivariate vector autoregression model, while varying the window length to validate the robustness of our results \cite{box_time_2008}.
Figure \ref{fig_sm_gi_window} displays the temporal evolution of the Granger index over four distinct windows, ranging from 13 to 17 days, and Figure \ref{fig_sm_gi_window} displays the coefficient of multivariate vector autoregression models over the same windows. 
Consistently, we observe results aligning with those presented in the main text, particularly in relation to the three phases.
We note that the Granger index diminishes as the window length increases, yet the ratio between Granger indices, or the ratio of a Granger index at one time to that at a subsequent time, remains constant. 
This variation is not observed in the coefficients of the Multivariate Vector Autoregression models, given that we are modeling log-returns.
However, we also note a shift (of one day) in the third phase in both these results.

\begin{figure}[h!]
    \centering
    \includegraphics[width=0.75\textwidth]{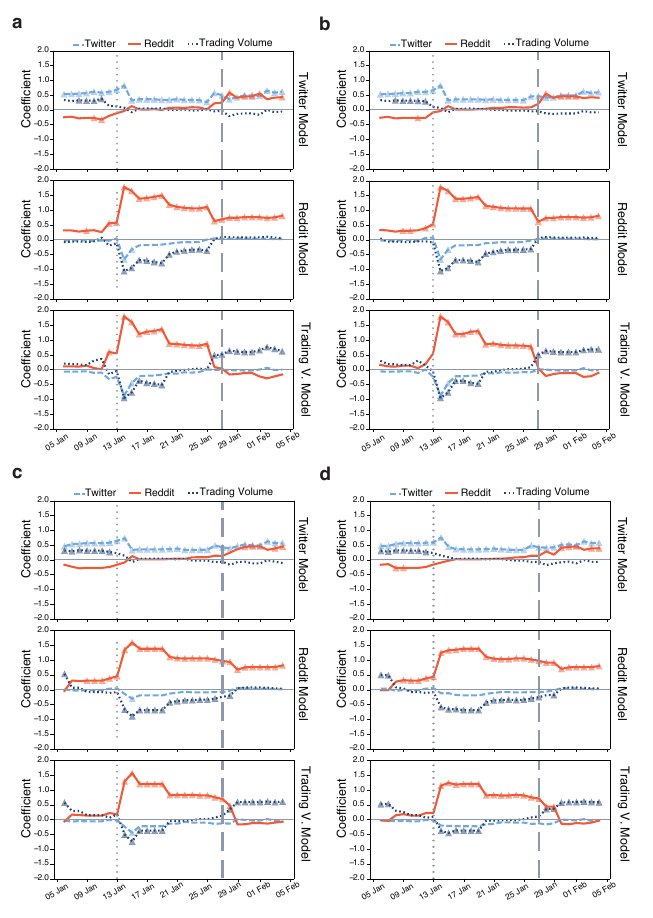}
    \caption{Each sub-panel shows the coefficients of a multivariate vector autoregressive model predicting Twitter (top), Reddit (middle), and Trading Volume (bottom) activities, from antecedent values of Reddit (solid orange), Twitter (dashed lightblue), and trading volume (dotted blue).
    Each point is computed considering time series spanning the $w$ preceding days.
    In panel \textbf{a} $w = 13$, in panel \textbf{b} $w = 14$, in panel \textbf{c} $w = 16$ and in panel \textbf{d} $w = 17$.
    The vertical gray dotted line indicates the initiation of the Action phase (13 January 2021), whereas the vertical gray dashed line corresponds to the tweet by Elon Musk (27 January 2021). 
    The triangles refer to statistical significant test (p-value $p<0.1$) within the window.
    }
    \label{fig_sm_mva_window}
\end{figure}

\pagebreak[2]
\clearpage

\subsection*{Granger Test and Multivariate Vector Autoregression model on daily log returns}

\begin{figure}[h!]
    \centering
    \includegraphics[width=1.0\textwidth]{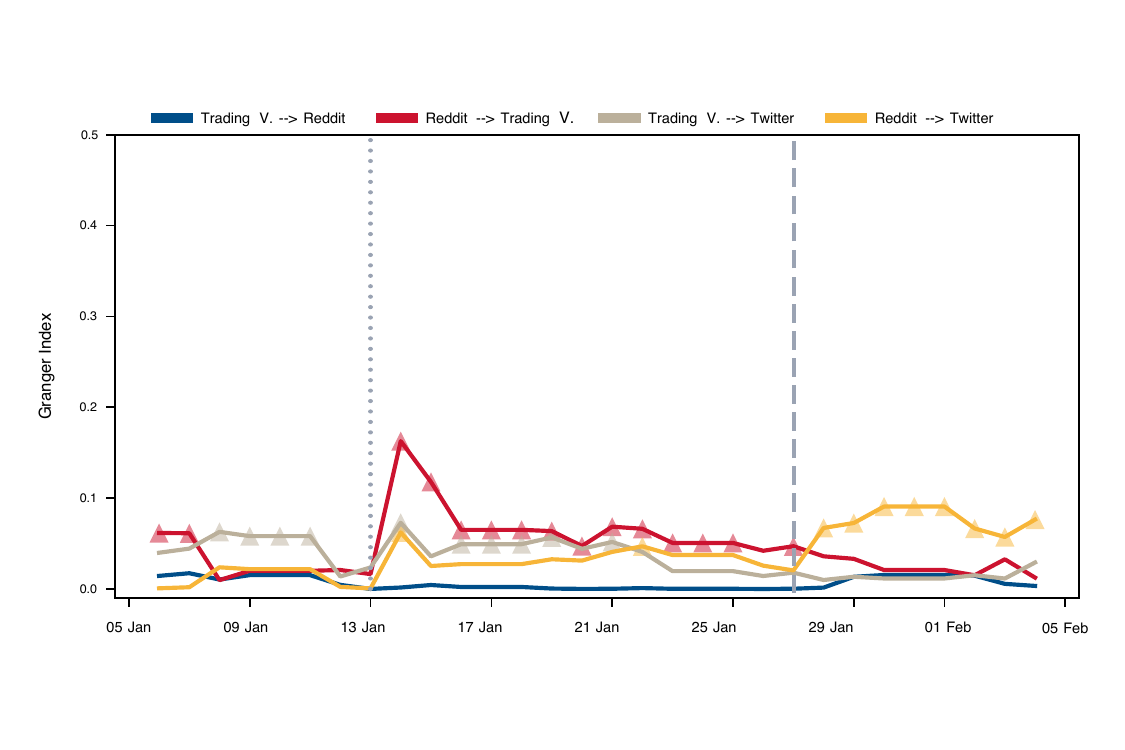}
    \caption{Granger Index capturing the predictive power of a signal on another (using daily log return with a lag of 1 hour) for the following pairs: Trading Volume-to-Reddit (blue), Reddit-to-Trading Volume (red), Trading Volume-to-Twitter (gray) and Reddit-to-Twitter (yellow).
    The time windows span 15-day, corresponding to 120 points per time series.
    The vertical gray dotted line indicates the initiation of the Action phase (13 January 2021), whereas the vertical gray dashed line corresponds to the Tweet by Elon Musk (27 January 2021) that marks the start of the \emph{Visibility} phase.
    The triangles correspond to statistical significant values (p-values $<0.1$).
    }
    \label{fig_sm_gi_smooth}
\end{figure}
\pagebreak[2]
\clearpage
\begin{figure}[b!]
    \centering
    \includegraphics[width=0.85\textwidth]{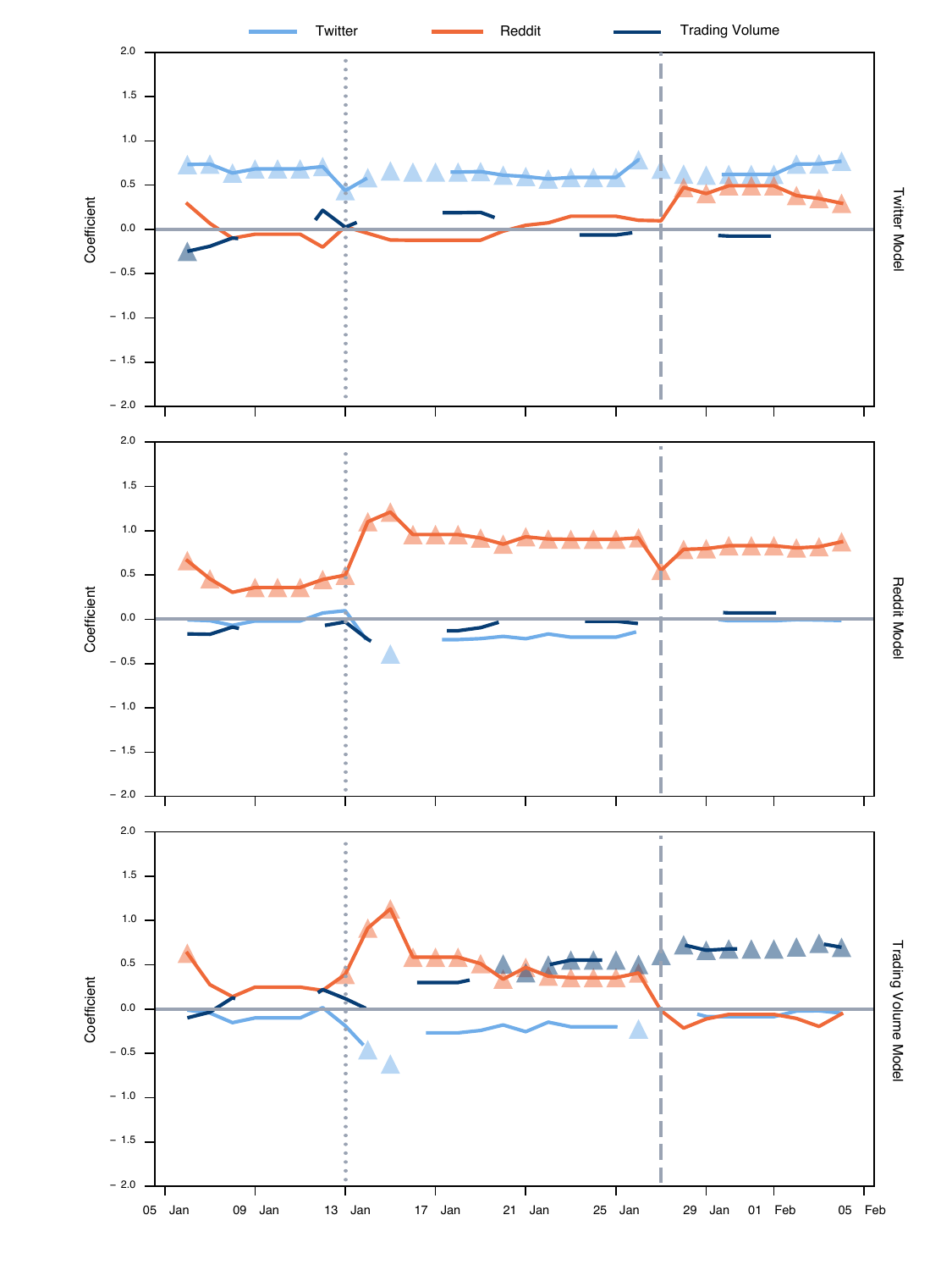}
    \caption{Each panel shows the coefficients of the multivariate autoregressive model using daily log return at lag 1 hour predicting Twitter (top), Reddit (middle), and Trading Volume (bottom) activities, from antecedent values of Reddit (orange), Twitter (lightblue), and Trading Volume (blue). 
    The time windows span 15-day, corresponding to 120 points per time series.
    The vertical gray dotted line indicates the initiation of the Action phase (13 January 2021), whereas the vertical gray dashed line corresponds to the tweet by Elon Musk (27 January 2021). 
    The triangles refer to statistical significant test (p-value $p<0.1$) within the 15-days window.
    }
    \label{fig_sm_coeff_smooth}
\end{figure}
\pagebreak[2]
\clearpage

\subsection*{Multivariate Vector Autoregression AMC and NOK during January 2021}
\begin{figure}[h!]
    \centering
    \includegraphics[width=1.0\textwidth]{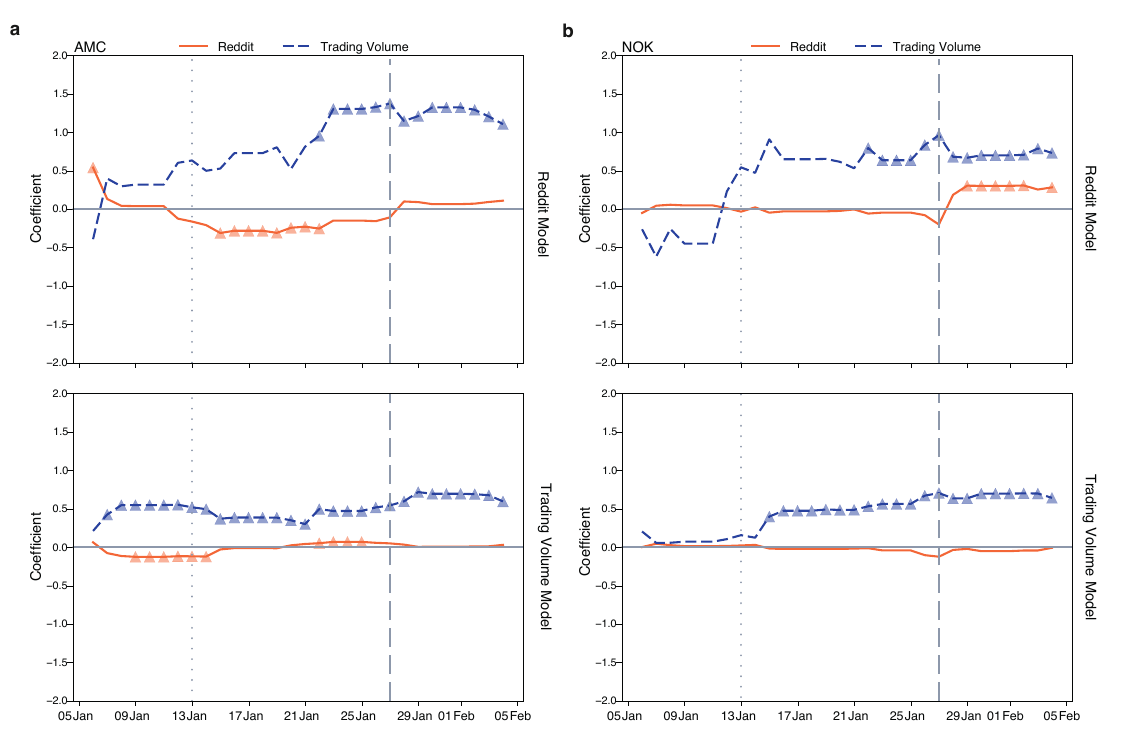}
    \caption{Each panel shows the coefficients of an autoregressive model predicting Reddit (top) and Trading Volume (bottom) activities, from antecedent values of Reddit (orange) and Trading Volume (blue). 
    The triangles correspond to statistical significant values (p-values $<0.1$).
    The x-values of the lineplots correspond to the last day of each 15-days window.
    In all the subplots the vertical gray dotted line indicates the initiation of the Action phase (13 January 2021), whereas the vertical gray dashed line corresponds to the Visibility phase (27 January 2021). 
    In panel \textbf{a} AMC and in panel \textbf{b} NOK.
    }
    \label{fig_sm_coeff_jan}
\end{figure}
\pagebreak[2]
\clearpage
\section*{Supplementary Materials, Section 6: Reply speed}\label{sm_users_stock}
\begin{figure}[h!]
    \centering
    \includegraphics[width=1.0\textwidth]{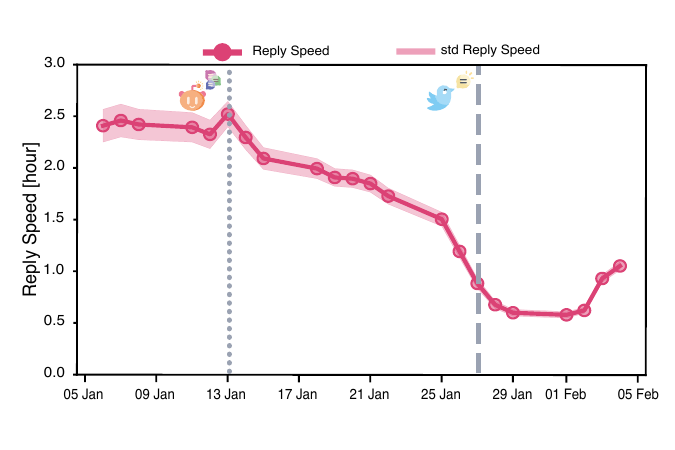}
    \caption{Daily average (and standard deviation of the mean) of the reply speed, defined as the average time interval between a comment and its direct responses under a Reddit post. The average by post is weighted using the number of comments underneath the post. We apply a 5-day moving average to the time series and consider only weekdays to enhance the chart's clarity. 
    }
    \label{fig_sm_reply_speed}
\end{figure}

\pagebreak[2]
\clearpage

\section*{Supplementary Materials, Section 7: Users' interests in stocks and shifts in January}\label{sm_users_stock}
\begin{figure}[h!]
    \centering
    \includegraphics[width=1.0\textwidth]{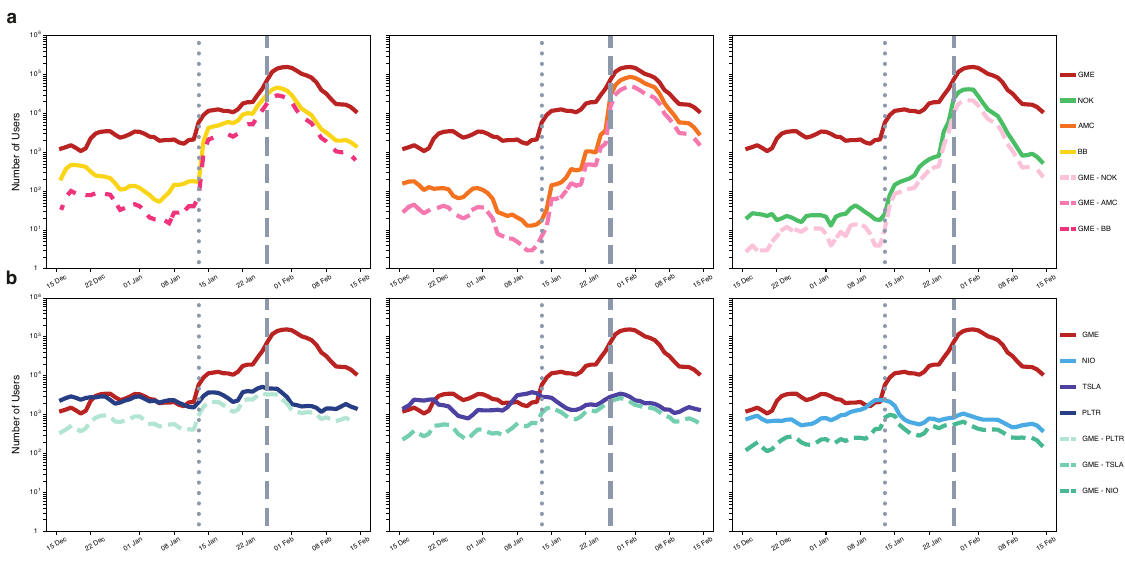}
    \caption{\textbf{Raw Number of users talking of stock.}
    Each subpanel represents the number of users discussing GME (red solid line), a stock and both (dashed lines).
    Panel \textbf{a} displays the stocks that gained popularity in January: BB (left panel), AMC (central panel) and NOK (right panel).
    Panel \textbf{b} displays the stocks that are no longer in the top rank: PLTR (left panel), TSLA (central panel) and NIO (right panel).
    The vertical gray dotted line indicates the beginning of the Action phase (13 January 2021), whereas the vertical gray dashed line corresponds to the Tweet by Elon Musk (27 January 2021).
    }
    \label{fig_sm_raw_users}
\end{figure}
\pagebreak[2]
\clearpage
\begin{figure}[!t]
    \centering
    \includegraphics[width=\textwidth]{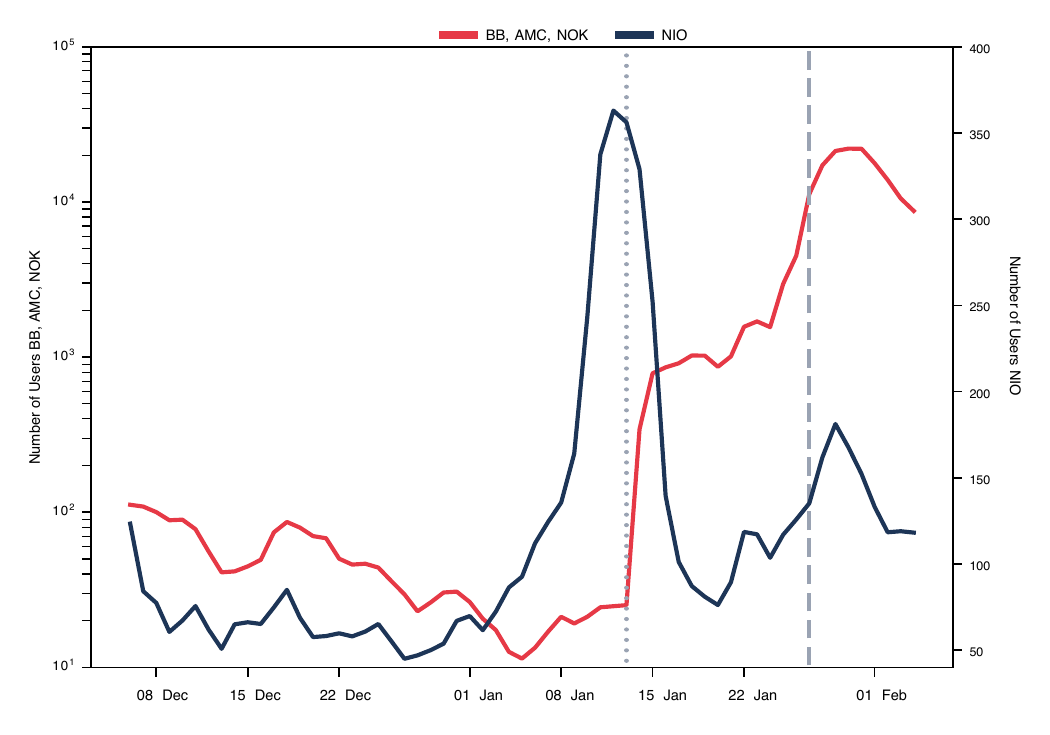}
    \caption{\textbf{Users shift from NIO to BB, AMC, NOK during January.}
    The plot displays the dynamic shift in user engagement, contrasting the raw number of users ``leaving'' NIO with the raw number of users ``adopting'' BB, AMC, NOK over time. 
    ``Leaving'' users are those who have stopped discussing NIO, while "adopting" users are those initiating discussions for the first time.
    The vertical gray dotted line indicates the beginning of the Action phase (13 January 2021), whereas the vertical gray dashed line corresponds to the Tweet by Elon Musk (27 January 2021).
    }
    \label{fig_sm_leaving_adopting}
\end{figure}
\end{document}